\documentclass[journal=jacsat,manuscript=article]{achemso}
\setkeys{acs}{articletitle = true}
\usepackage[version=3]{mhchem} 
\usepackage[usenames,dvipsnames]{xcolor}
\usepackage{textcomp,amssymb,float} 

\renewcommand{\vec}[1]{\boldsymbol{\mathrm{#1}}}
\newcommand{\uvec}[1]{\hat{\mathbf{#1}}}%
\newcommand{\fct}[1]{\operatorname{#1}}%
\newcommand{\abs}[1]{\lvert#1\rvert}%
\newcommand{\mean}[1]{\left\langle #1 \right\rangle}%
\newcommand{\DiracDelta}[1]{\operatorname{\delta}\left(#1\right)}%

\renewcommand{\bar}[1]{\overline{#1}}%

\newcommand{\dif}{d}%
\newcommand{\del}[1]{\delta \hspace{-2pt} \fct{#1}}%
\newcommand{\Del}[1]{\Delta \hspace{-1pt} #1}%

\newcommand{\integral}[4]{\int_{#1}^{#2} \! #3 \, \dif#4}%
\newcommand{\integralvar}[3]{\int_{#1}^{#2} \! \! \dif#3 \!}%
\newcommand{\ii}{\mathrm{i}}

\author{Alexander R. Sprenger}
\affiliation[Heinrich-Heine-Universit\"at D\"usseldorf]{Institut f\"ur Theoretische Physik II: Weiche Materie, Heinrich-Heine-Universit\"at D\"usseldorf, D-40225 D\"usseldorf, Germany}

\author{Miguel Angel Fernandez-Rodriguez}
\affiliation[ETH Z\"urich]{Laboratory for Soft Materials and Interfaces, Department of Materials, ETH Z\"urich, 8093 Z\"urich, Switzerland}

\author{Laura Alvarez}
\affiliation[ETH Z\"urich]{Laboratory for Soft Materials and Interfaces, Department of Materials, ETH Z\"urich, 8093 Z\"urich, Switzerland}

\author{Lucio Isa}
\affiliation[ETH Z\"urich]{Laboratory for Soft Materials and Interfaces, Department of Materials, ETH Z\"urich, 8093 Z\"urich, Switzerland}
\email{lucio.isa@mat.ethz.ch}

\author{Raphael Wittkowski}
\affiliation[Westf\"alische Wilhelms-Universit\"at M\"unster]{Institut f\"ur Theoretische Physik, Center for Soft Nanoscience, Westf\"alische Wilhelms-Universit\"at M\"unster, D-48149 M\"unster, Germany}
\email{raphael.wittkowski@uni-muenster.de}

\author{Hartmut L\"owen}
\affiliation[Heinrich-Heine-Universit\"at D\"usseldorf]{Institut f\"ur Theoretische Physik II: Weiche Materie, Heinrich-Heine-Universit\"at D\"usseldorf, D-40225 D\"usseldorf, Germany}

\title{Active Brownian motion with orientation-dependent motility: theory and experiments} 

\abbreviations{}
\keywords{}

\begin{document}

\begin{tocentry}
\begin{center}
\includegraphics[width=6.5cm]{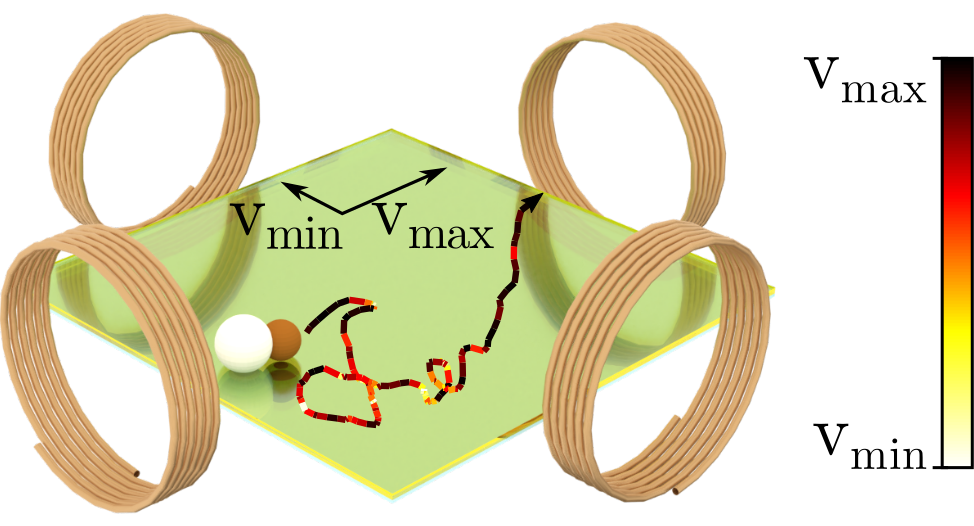}
\end{center}
\end{tocentry}

\pagebreak
\begin{abstract}
Combining experiments on active colloids, whose propulsion velocity can be controlled via a feedback loop, and theory of active Brownian motion, we explore the dynamics of an overdamped active particle with a motility that depends explicitly on the particle orientation. In this case, the active particle moves faster when oriented along one direction and slower when oriented along another, leading to an anisotropic translational dynamics which is coupled to the particle's rotational diffusion. We propose a basic model of active Brownian motion for orientation-dependent motility. Based on this model, we obtain analytic results for the mean trajectories, averaged over the Brownian noise for various initial configurations, and for the mean-square displacements including their anisotropic non-Gaussian behavior. The theoretical results are found to be in good agreement with the experimental data. Our findings establish a methodology to engineer complex anisotropic motilities of active Brownian particles, with potential impact in the study of the swimming behavior of microorganisms subjected to anisotropic driving fields.
\end{abstract}

\pagebreak
\section{Introduction}
Active Brownian particles, the synthetic analogues of biological microswimmers such as bacteria and protozoa, have the ability to self-propel at low Reynolds numbers via the conversion of energy available in their surroundings into directed motion, by exploiting intrinsic asymmetries in their shape and material properties \cite{BechingerdLLRVV2016, Howse2010}. Their motion arises from the interplay between thermal fluctuations and propulsion, which renders active colloids an excellent model system to study far-from-equilibrium physical phenomena \cite{RomanczukBELSG2012,ElgetiWG2015,ZoettlS2016}, also featured in their biological counterparts. The basic model to describe the trajectories of a self-propelling colloid, called active Brownian motion, couples a constant velocity $v$ along the particle's asymmetry direction with its rotational diffusivity $D_\text{R}$, which constantly randomizes the propulsion direction with a characteristic time scale $\tau_\text{R} = 1/D_\text{R}$. In this model, the particle displacements result from propulsion combined with stochastic translational and rotational noise. The propensity for straight paths is defined by the persistence length of the trajectory $L_\text{P}= v /D_\text{R}$. To date, various propulsion mechanisms have been realized for active colloids. Among them are self-propulsion induced by chemical reactions \cite{PaxtonKOSSCMLC2004,PalacciCYB2010,DietrichRZVBI2017}, illumination \cite{VolpeBVKB2011,ButtinoniVKVB2012,PalacciSSPC2013,PalacciSVCP2013,PalacciSKYPC2014,MoysesPSG2016}, or ultrasound \cite{WangCHM2012} and actuation by magnetic \cite{DreyfusBRFSB2005,GrosjeanLDHLV2015,SteinbachGE2016,KaiserSA2017} or electric \cite{BricardCDDB2013,MorinDCB2017} fields. Regardless of the origin of propulsion, the scenario defined by active Brownian motion \cite{BechingerdLLRVV2016} was verified in experiments for a range of artificial microswimmers \cite{KuemmeltHWBEVLB2013,KuemmeltHWTBEVLB2014,ZhengtHKWCSLL2013}.

Despite the success of ordinary active Brownian motion, the complexity of some behaviors found in biological and artificial microswimmers implies the urge to extend our experimental and theoretical models, in particular to include complex spatio-temporal dependencies of propulsion velocity as well as translational and rotational noise. These situations are frequently encountered for systems where the external stimulus governing the motility is inhomogeneous. \cite{HongBKSV2007,PohlS2014,SahaGR2014,LiebchenMPC2015,LiebchenMC2017,JinKM2017,LozanotHLB2016,GarciaRP2013}. Recently, motility landscapes, where the particle propulsion speed depends on spatial coordinates, time, or a combination of both \cite{GeiselerHMMS2016,GeiselerHM2017a,GeiselerHM2017b,SharmaB2017}, have been experimentally realized \cite{HongBKSV2007,LozanotHLB2016,DaiWXZDLFT2016,LiWQZL2016,LozanoB2019,LozanoLtHBL2019,PalagiSF2019} and numerically modeled \cite{GhoshLMN2015,MagieraB2015,LozanotHLB2016,GrauerLJ2017,ScacchiBS2019,VuijkBS2019,MerlitzVBSS2018,LiebchenL2019,MaggiAFDiL2018}. 
However, with rare recent exceptions aside \cite{Uspal2019}, the orientational analogue to a position-dependent motility landscape, which is an orientation-dependent motility, remains unexplored for systems of non-interacting anisotropic active particles. 

In this article, we experimentally and theoretically study active dumbbells with an orientation-dependent motility. This system offers a basic set-up for anisotropic actuation, in which the particle's propulsion speed is modulated according to its orientation, which is constantly randomized by rotational diffusion, thus introducing an anisotropy in the particle dynamics. In our experiments, we use active dumbbell-shaped colloids composed of a polystyrene and a magnetic silica particle assembled via sequential capillary assembly \cite{NiLBIW2016} and self-propelling on a planar substrate via alternating electric fields \cite{EHDflow, NiMBWI2017}. The particle's position and orientation are tracked in real time and used as the input for a feedback loop that updates the particle velocity with full programmability \cite{Arxive_magnetic}. These results are used to verify the basic theoretical model for active Brownian motion with an orientation-dependent velocity, which we propose and establish here. We obtain analytic results for mean trajectories averaged over the Brownian noise for various initial configurations and arbitrary angular dependencies of the velocity. We further calculate the corresponding mean-square displacements, including their anisotropic non-Gaussian behavior, and characterize the anisotropy as a function of time. We find that the theoretical calculations are in good agreement with the experimental data. The results of this work shed new light on anisotropic active Brownian particles, inspiring both a better understanding of the behavior exhibited by motile microorganisms when subjected to inhomogeneous or anisotropic driving fields \cite{HuWWG2015} and new design ideas for smarter synthetic microswimmers.

\pagebreak
\section{\label{II}Materials and Methods} 
\subsection{Theoretical description} 
In our theoretical model, we consider a single overdamped active Brownian particle in two spatial dimensions. The state of this particle is fully described by the center-of-mass position $\vec{r}(t)$ and the angle of orientation $\phi(t)$, which denotes the angle between the orientation vector $\uvec{u} = (\cos\phi,\sin\phi)$ and the positive $x$-axis, at the corresponding time $t$. The centerpiece of our model is an arbitrary orientation-dependent motility $\vec{v}(\phi)$. Without loss of generality, we represent the propulsion velocity $\vec{v}(\phi)$ as a Fourier series
\begin{equation}%
\vec{v}(\phi) =  \bar{v} \sum_{k=-\infty}^{\infty} \vec{c}_{k} \exp ( \ii k \phi ),
\label{eq_velocity}%
\end{equation}%
where $\bar{v}$ denotes a reference velocity, $\vec{c}_{k}$ is the Fourier-coefficient vector of the mode $k$, and $\ii$ denotes the imaginary unit. For a given propulsion velocity $\vec{v}(\phi)$, these Fourier coefficients can be calculated as $\vec{c}_{k} =  \integral{-\pi}{\pi}{(\vec{v}(\phi)/(2 \pi \bar{v})) \exp(- \ii k \phi)}{\phi}$. The overdamped Brownian dynamics of the particle is described by the coupled Langevin equations for orientation-dependent motility
{\allowdisplaybreaks
\begin{align}%
\dot{\vec{r}}(t) & = \vec{v}\big( \phi(t) \big) + \sqrt{2 D_{ \text{T} } } \vec{\xi}(t),  
\label{eq_langevin_trans}\\%
\dot{\phi}(t)    & =  \sqrt{2 D_{\text{R}} } \eta(t),
\label{eq_langevin_rot}%
\end{align}}%
where $D_{\text{T}}$ and $D_{\text{R}}$ are the translational and rotational short-time diffusion coefficients of the particle, respectively. To take translational and rotational diffusion into account, the Langevin equations contain independent Gaussian white noise terms $\vec{\xi}(t)$ and $\eta(t)$, with zero means $ \mean{\vec{\xi}(t)} = \vec{0}$ and $ \mean{\eta(t)} = 0$ and delta-correlated variances $\mean{ \xi_{i} (t_{1}) \xi_{j}(t_{2})} = \delta_{ij} \DiracDelta{t_{1} - t_{2} }$ and $\mean{ \eta(t_{1}) \eta(t_{2})} = \DiracDelta{t_{1} - t_{2} } $, where $i, j \in \{x, y\}$. The brackets $\mean{\dots}$ denote the noise average and $\delta_{ij}$ is the Kronecker delta. Note that we describe the self-propulsion with the prescribed vector function by $\vec{v}(\phi)$ (and not with a scalar prefactor by $\fct{v}_{0}(\phi) \uvec{u}(\phi)$ as typically assumed for isotropic self-propulsion \cite{BechingerdLLRVV2016}). In the following, we neglect the mode $k=0$ in Eq.~\eqref{eq_velocity}, which would describe a trivial constant drift. 

\subsection{Fabrication of active magnetic dumbbells}
Active magnetic dumbbells composed of a 2.0 $\mu \text{m}$-diameter polysterene (PS) and a 1.7 $\mu \text{m}$-diameter magnetic silica (\ce{SiO2}-mag) particle (Microparticles GmbH) were fabricated using the sequential Capillarity-Assisted Particle Assembly (sCAPA) technique as described in previous work \cite{NiLBIW2016}. First, a 40 $\mu \text{l}$ water droplet (Milli-Q) with 0.1 mM sodium dodecyl sulfate (SDS, 99.0 \%, Sigma-Aldrich), 0.01 \%wt of the surfactant Triton X-45 (Sigma), and 0.5 \%wt PS particles was deposited and dragged at a controlled speed over a polydimethylsiloxan (PDMS) template with rectangular traps of 2.2 $\mu \text{m}$ $\times$ 1.1 $\mu \text{m}$ lateral dimensions and 0.5 $\mu \text{m}$ depth, fabricated by conventional photolithography. This deposition step resulted in one PS particle deposited per trap, leaving space for a second particle. The process was then repeated with a dispersion of \ce{SiO2}-mag particles. Individual \ce{SiO2}-mag particles were deposited inside the traps in close contact with the PS particles forming dumbbells. Next, the dumbbells were sintered in the traps by heating the template to 85 \textdegree C for 25 minutes. Finally, the dumbbells were harvested by freezing a droplet of a 10 $\mu \text{M}$ \ce{KCl} (Fluka) aqueous solution over the traps and lifting it from the template. The thawed droplet containing the dumbbells was used to fill the experimental cell as described below. 

\subsection{Cell preparation and active motion control}
Transparent electrodes were fabricated from 22 mm $\times$ 22 mm glass slides ($85-115\,\mu\text{m}$ thick, Menzel Gl\"aser, Germany) coated via e-beam metal evaporation with 3 nm \ce{Cr} and 10 nm \ce{Au} (Evatec BAK501 LL, Switzerland), followed by a top layer of 10 nm of \ce{SiO2} (STS Multiplex CVD, UK) deposited by plasma-enhanced chemical vapor deposition. A 7.4 $\mu \text{l}$ droplet of the dumbbells suspension was placed on the bottom electrode inside a 0.12 mm-thick sealing spacer with a 9 mm-circular opening (Grace Bio-Labs SecureSeal, USA). 

After sealing the cell with the top electrode, both electrodes were connected to a signal generator (National Instruments Agilent 3352X, USA) to apply an AC electric field with a fixed frequency of 1 kHz and varying peak-to-peak voltage $V_\text{PP}(t)$ between 1 and 10 V, depending on the dumbbell orientation. The particles propel thanks to unbalanced electrohydrodynamic (EHD) flows on each side of the dumbbell, with the \ce{SiO2}-mag lobe leading the motion. The propulsion velocity is proportional to ${V_\text{PP}}^2$ \cite{EHDflow, NiMBWI2017}.

We furthermore imposed a fixed rotational diffusivity $D_{ \text{R} }=0.25\,\text{rad}^2/\text{s}$ of the dumbbells for all experiments, as described in a previous work \cite{Arxive_magnetic}. In brief, we applied external magnetic fields via two pairs of independent Helmholtz coils to align the magnetic moment of the \ce{SiO2}-mag particle. The angle $\phi(t)$ of the applied magnetic field is randomly varied in time according to the relation $\phi(t+\Delta t)=\phi(t)+\sqrt{2 D_{\text{R}} \Delta t} \, \eta(t)$, where in the experiments $\Delta t=1\,\text{ms}$ and $\eta(t)$ is defined as above.

\subsection{Imaging and feedback loop}
The dumbbells were imaged in transmission with a home-built bright-field microscope. Image sequences were taken with a sCMOS camera (Andor Zyla) with a 1024 pixels $\times$ 1024 pixels field of view and a $50\times$ objective (Thorlabs). The center of mass $\vec{r}(t)$ and the angle $\phi(t)$ of the dumbbells with respect to the $x$-axis were tracked in real time using a customized software written in Labview and Matlab. The detected orientation of the dumbbell is symmetric with respect to $\pi$, being 0 or $\pi$ when it is perfectly aligned with the $x$-axis. After the experiments, we post-processed the acquired images to identify both lobes of the dumbbell and convert the angles to the interval from 0 to $2\pi$. The velocity of the dumbbell was varied as a function of its orientation by changing the applied peak-to-peak voltage $V_\text{PP}$ according to
\begin{equation}
V_\text{PP}(t)= (V_\text{PP}^\text{max}-V_\text{PP}^\text{min}) \sin^{2}\big(n \phi(t)\big) + V_\text{PP}^\text{min},
\label{eq:theta2}%
\end{equation}
where $V_\text{PP}^\text{max}$ and $V_\text{PP}^\text{min}$ are the maximum and minimum values of the applied peak-to-peak voltage and $n=1,2$ is the number of symmetric lobes in $\vec{v}(\phi)$. For $n=1$, the dumbbell velocity is maximal when the particle is aligned with the $y$-axis and minimal when it is aligned with the $x$-axis. In the case of $n=2$, the dumbbell velocity is maximal for an orientation angle $\pi/4$ and minimal when the particle is aligned with the $x$- or $y$-axis.

There is an inherent delay between capturing an image, extracting the dumbbell angle, and updating the voltage according to it. In our experimental setup, a full cycle takes 400 ms, leading to an update frequency of the particle velocity of 2.5 Hz. This frequency is much lower than the one used to randomize the dumbbell orientation (1 kHz), so that there is a clear separation of time scales between the two types of updates and the dumbbell undergoes standard rotational diffusion at an imposed rate.

\pagebreak
\section{\label{III}Results and discussion} 
\subsection{Orientation-dependent motility}

\begin{figure}[H]
\centering
\includegraphics[width=0.8\columnwidth]{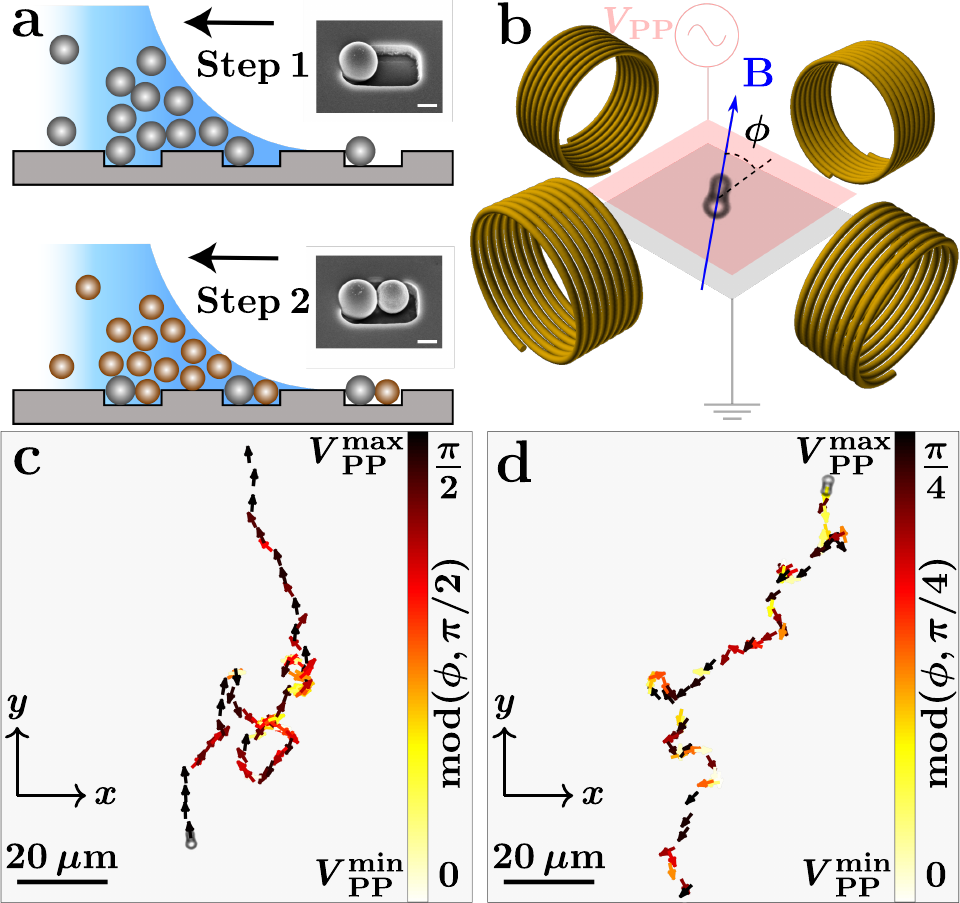}
\caption{\label{fig1}\textbf{a} Side-view representation of the sCAPA fabrication of active magnetic dumbbells. The PS particles (gray spheres) are deposited first, followed by the \ce{SiO2}-mag particles (brown spheres). The black arrows indicate the deposition direction. The insets show SEM images of the particles in the traps after each deposition step (2 $\mu$m-scale bar). \textbf{b} Scheme of the experimental setup. Four magnetic coils impose a randomly oriented magnetic field $\vec{B}$ (blue arrow) to set the rotational diffusivity of the dumbbells to $D_\text{R}=0.25\,\text{rad}^2/\text{s}$. An AC electric field applied between two transparent electrodes is used to actuate the dumbbell with velocity $v \propto V_\text{PP}^2(t)$. A feedback loop updates the applied voltage as a function of the dumbbell orientation angle $\phi(t)$ to achieve an orientation-dependent propulsion velocity. \textbf{c-d} Trajectories of active magnetic dumbbells with a motility with two-fold (c) and four-fold (d) rotational symmetry. The particle positions at discrete times are represented by arrows indicating the dumbbell orientation and color-coded according to the applied voltage in the range from $V_\text{PP}^\text{min}$ to $V_\text{PP}^\text{max}$, which corresponds to $\fct{mod}(\phi(t),\pi/2)$ (c) and $\fct{mod}(\phi(t),\pi/4)$ (d). See the corresponding Supplementary Movies.}
\end{figure}

Our active colloidal dumbbells are produced by sequential capillary assembly \cite{NiLBIW2016}, as represented in Fig.~\ref{fig1}a in Materials and Methods, and self-propel under AC electric fields thanks to induced-charge electrophoresis \cite{squires_bazant_2006,Gangwal2008,Yan2016}. The compositional asymmetry of the dumbbell results in local unbalanced EHD flows producing a net force that generates propulsion along the long axis of the dumbbell \cite{EHDflow, NiMBWI2017}. In order to achieve robust experimental control of orientational dynamics, we decouple it from the thermal bath by randomizing the dumbbell orientation using an external magnetic field (see Fig.~\ref{fig1}b) to set a constant rotational diffusivity of $D_\text{R}=0.25\,\text{rad}^2/\text{s}$ \cite{Arxive_magnetic}. We furthermore include a feedback loop to update the dumbbell's propulsion velocity according to its orientation, as described in Materials and Methods and sketched in Fig.~A1 in the Appendix, to experimentally realize active Brownian particles with orientation-dependent motility. 

In this work, we study two representative orientation-dependent motilities. In the first case, the particle's motility has a two-fold rotational symmetry, with the lowest velocity when the particle is oriented along the $x$-axis and the highest when it is oriented along the $y$-axis (see Fig.~\ref{fig1}c and Supplementary Movie 1). We incorporate this motility effectively in leading order as
\begin{equation}%
\vec{v}_{1}(\phi) =  2 \, \bar{v}_{1} \sin^{2} \phi \, \uvec{u}(\phi), 
\label{eq_velocity_1}%
\end{equation}%
where $\bar{v}_{1}$ denotes the orientationally averaged speed of the particle. In the second case, the velocity has a four-fold symmetry, where the dumbbell achieves the highest velocity when it is aligned along the diagonal corresponding to an orientation angle $\pi/4$ and the lowest when it is aligned with the $x$- or $y$-axis (see Fig.~\ref{fig1}d and Supplementary Movie 2). This case is analogously described as 
\begin{equation}%
\vec{v}_{2}(\phi) =  2 \, \bar{v}_{2} \sin^{2}( 2 \phi ) \uvec{u}(\phi).
\label{eq_velocity_2}%
\end{equation}%

\begin{figure*}[h]
\centering
\includegraphics[width=8cm]{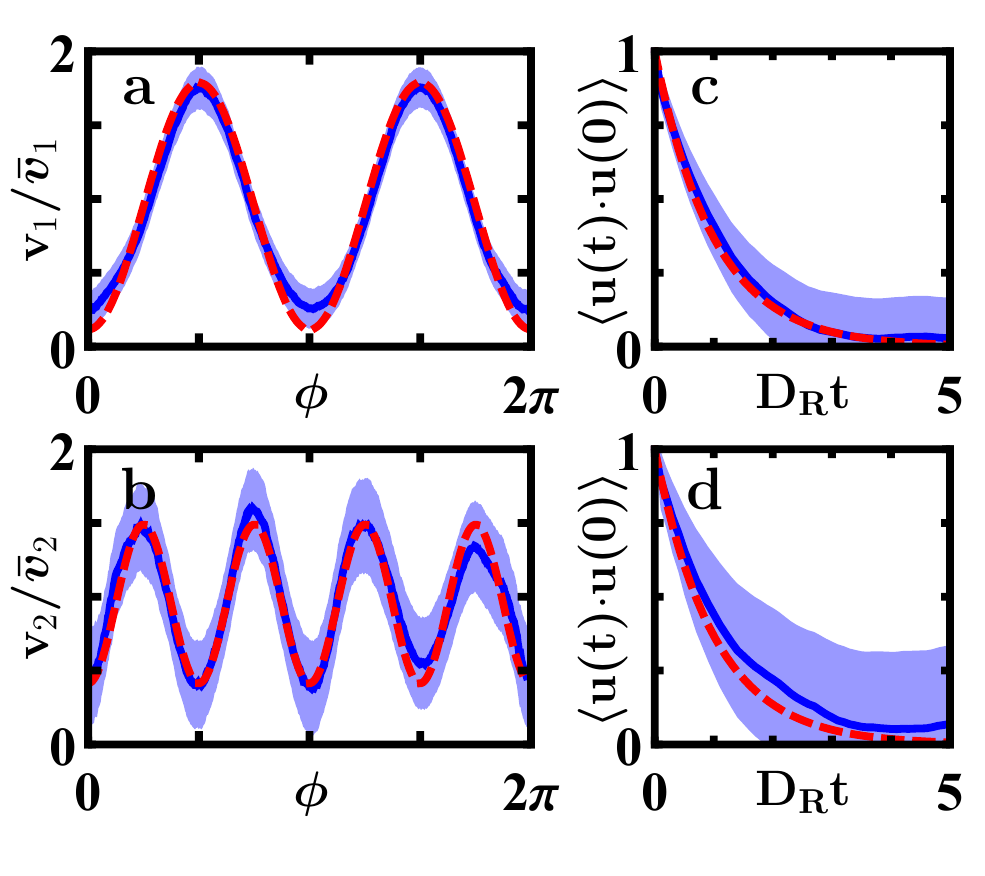}
\caption{\label{fig2}\textbf{a}-\textbf{b} Orientation-dependent motility with two-fold rotational symmetry $\fct{v}_{1}(\phi)/\bar{v}_{1} =  2 \sin^{2} \phi $ and four-fold rotational symmetry $\fct{v}_{2}(\phi)/\bar{v}_{2} =  2 \sin^{2}( 2 \phi )$. Solid dark blue and dashed red curves show the experimental data and a trigonometric fit, respectively. The fits yield $\bar{v}_{1} =1.4 \, \mu \text{m}/\text{s}$ and $\bar{v}_{2} =1.1 \, \mu\text{m}/\text{s}$. Light blue areas express the standard error of the mean. \textbf{c}-\textbf{d} Orientation-correlation function $\mean{\uvec{u}(t) \cdot \uvec{u}(0)}$ for the two experiments and the expected function $\mean{\uvec{u}(t) \cdot \uvec{u}(0)} = \exp( - D_{\text{R}} t )$ for comparison, validating the imposed rotational diffusivity $D_{\text{R}} =0.25 \, \text{rad}^2/\text{s}$.}
\end{figure*}

Figure~\ref{fig2} shows that the prescribed motility scenarios are experimentally realized. In Fig.~\ref{fig2}a-b we fit Eqs.~\eqref{eq_velocity_1} and \eqref{eq_velocity_2} to the data for the orientation-dependent velocity observed in the experiments corresponding to the first and second scenario, respectively. We find good agreement of the fit curves and experimental data and determine the orientationally averaged speeds $\bar{v}_{1}=1.4\,\mu \text{m}/\text{s}$ and $\bar{v}_{2} = 1.1\,\mu\text{m}/\text{s}$. The orientational decorrelation of the velocity vector obeys a simple exponential decay with a rate corresponding to the imposed rotational diffusivity $D_{ \text{R} } = 0.25\,\text{rad}^2/\text{s}$ (see Fig.~\ref{fig2}c-d). In the following, we will denote all lengths in units of the orientationally averaged persistence length $L  = \bar{v} / D_{\text{R}}$ (i.e., $\vec{r} \to \vec{r} / L$) and time in units of the persistence time $\tau_{ \text{R} } = 1/D_R$ (i.e., $t \to D_{\text{R}} t$). The importance of translational noise relative to the imposed speed $\bar{v}$ and rotational diffusion can be defined by the dimensionless P\'{e}clet number $\text{Pe} = \bar{v} / \sqrt{ D_{\text{R}} D_{\text{T}} }$, where the thermal translational diffusion coefficient of the dumbbells was experimentally determined to be $D_{\text{T}} = 0.055 \,\mu\text{m}^{2}/\text{s}$. 

\subsection{Mean displacement}
\begin{figure*}[h]
\centering
\includegraphics[width=16cm]{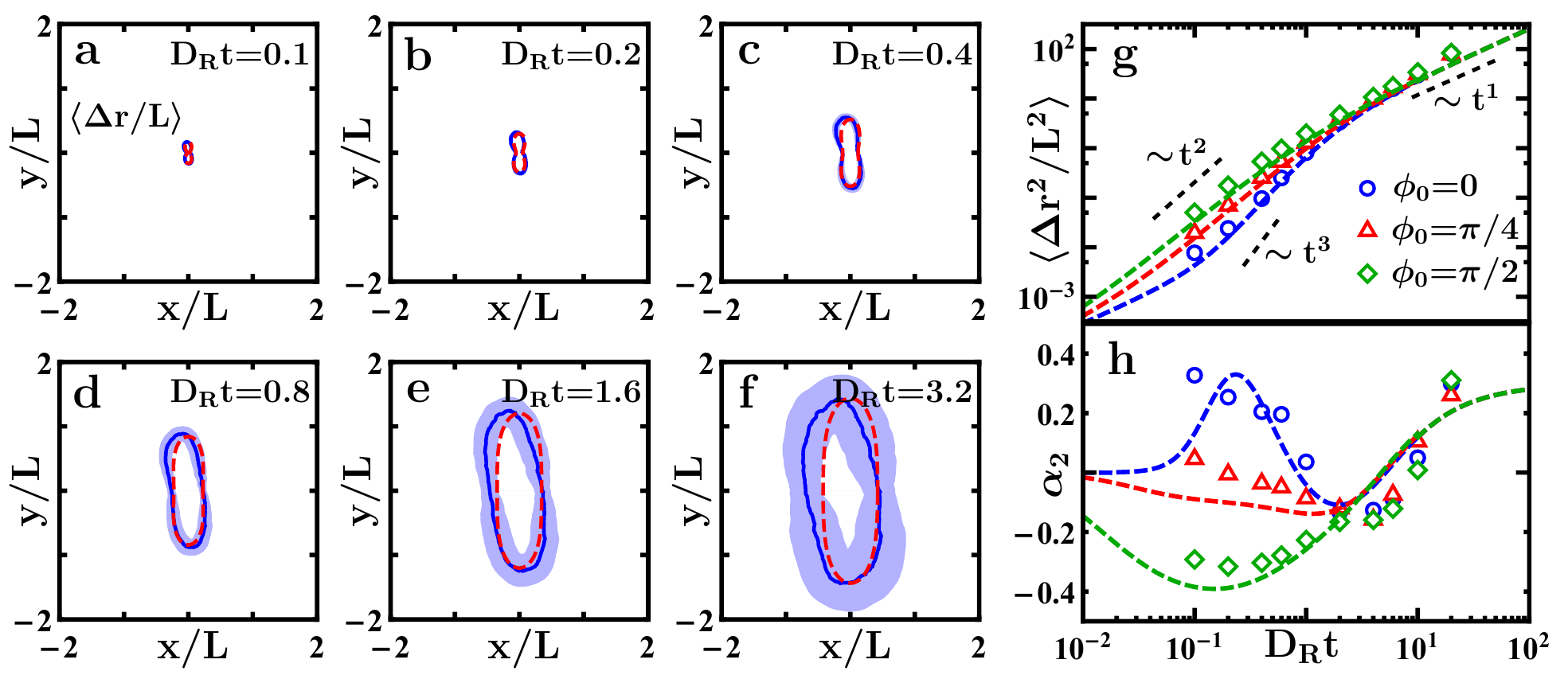}
\caption{\label{fig3}Comparison between theoretical and experimental results for a propulsion velocity with two-fold symmetry. \textbf{a}-\textbf{f} The anisotropic motion of the particle is visualized by plotting the mean displacement $\mean{\Delta \vec{r}(\phi_{0})}$ as a function of the initial orientation $\phi_{0}$ for fixed times $D_{ \text{R} } t=0.1$ (a), $D_{ \text{R} } t=0.2$ (b), $D_{ \text{R} } t=0.4$ (c), $D_{ \text{R} } t=0.8$ (d), $D_{ \text{R} } t=1.6$ (e), and $D_{ \text{R} } t=3.2$ (f). Solid dark blue and dashed red curves show the experimental data and analytic results, respectively. Light blue areas express the standard error of the mean. \textbf{g} The mean-square displacement $\mean{\Del{\vec{r}}^2(t)}$ for initial orientations $\phi_{0}=0$ (blue), $\phi_{0}=\pi/4$ (red), and $\phi_{0}=\pi/2$ (green). Symbols and dashed curves show the experimental data and analytic results, respectively. \textbf{h} The non-Gaussian parameter $\alpha_{2}(t)$ for the same initial orientations. Lengths are given in units of $L = 5.6\, \mu \text{m}$, time in units of $1/D_{\text{R}} = 0.4\,\text{s}$, and the P\'{e}clet number is set to $\text{Pe} = 12$.}
\end{figure*}

To characterize the effect of an orientation-dependent motility on the Brownian dynamics, we first discuss the mean displacement $\mean{\Del{\vec{r}}(t)}$ of the particle. In Figs.~\ref{fig3}a-f and \ref{fig4}a-f, the experimentally determined mean displacement is compared with the one resulting from our theoretical model, where we emphasize the anisotropic motion of the particle by plotting the mean displacement as a function of the initial orientation $\phi_{0} = \phi(0)$ after fixed times $t$. The theoretical result for the mean displacement is given for a general orientation-dependent motility as
\begin{equation}%
\frac{\mean{ \Del{\vec{r}}(t) }}{ L } = \sum_{\substack{k_{1}=-\infty \\ k_{1}\neq 0}}^{\infty}  \vec{c}_{k_{1}} \fct{C}_{k_{1}} (D_{\text{R}} t) \, e^{\ii k_{1} \phi_{0}}%
\label{eq_mean_displacement}%
\end{equation}%
with
\begin{equation}%
\fct{C}_{k_{1}}(t) = \frac{1}{k_{1}^{\,2}} \left( 1 - e^{-k_{1}^{\,2}t} \right), %
\end{equation}%
where the Fourier-coefficient vectors $\vec{c}_{k}$ are determined by the motility $\vec{v}(\phi)$, here $\vec{c}_k = \integral{-\pi}{\pi}{(\vec{v}_{n}(\phi)/(2 \pi \bar{v}_{n})) \exp(- \ii k \phi)}{\phi}$ for $n=1,2$. (All analytic results for the two studied scenarios are listed explicitly in the Appendix.) For short times $t \lesssim \tau_{ \text{R} }$, the particle moves linearly in time with $\mean{ \Del{\vec{r}}(t)  } = \vec{v}(\phi_{0}) t + \mathcal{O}(t^2)$ and the anisotropy with respect to the initial orientation, as is visible in Figs.~\ref{fig3}a and \ref{fig4}a, is a deterministic consequence of the anisotropic propulsion of the particle. For intermediate times $t \sim \tau_{ \text{R} }$, the orientation of the particle starts to decorrelate, which affects directly the anisotropic shape of the mean displacement (see Figs.~\ref{fig3}b-e and \ref{fig4}b-e). Finally, for long times $t \gtrsim \tau_{ \text{R} }$, the mean displacement saturates to an anisotropic persistence length $\lim\limits_{t \to \infty}  \mean{ \Del{\vec{r}}(t)  }  = L \sum_{k=-\infty,\,k\neq 0}^{\infty} \vec{c}_{k} e^{\ii k \phi_{0}} / k^2$ (see Figs.~\ref{fig3}f and \ref{fig4}f). The faster varying contributions, i.e., the higher Fourier modes, of the propulsion velocity saturate faster and have a smaller impact on the mean motion of the particle, resulting in a more isotropic final shape (compare Figs.~\ref{fig3}f and \ref{fig4}f).

\subsection{Mean-square displacement}
\begin{figure*}[h]
\centering
\includegraphics[width=16cm]{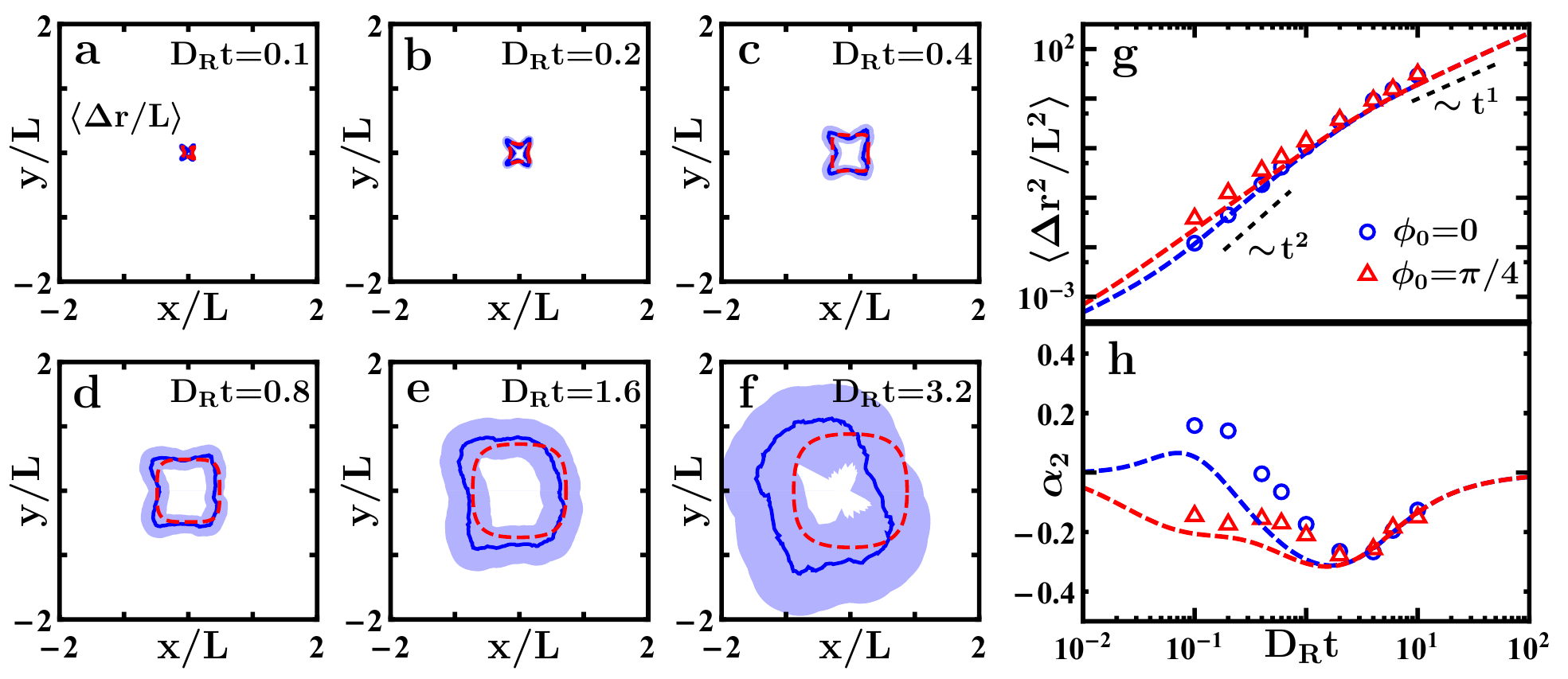}
\caption{\label{fig4}The same as in Fig.~\ref{fig3} for a propulsion velocity with four-fold symmetry. Lengths are given in units of $L = 4.4\, \mu \text{m}$, time in units of $1/D_{\text{R}} =0.4 \, \text{s}$, and the P\'{e}clet number is $\text{Pe} = 9$.}
\end{figure*}

The dynamics of active Brownian motion can be further classified in temporal regimes by investigating the scaling behavior of the mean-square displacement, i.e., $\mean{\Del{\vec{r}}^{2}(t)} \propto t^\nu$. For $\nu = 1$, the particle shows ordinary diffusive behavior. If $\nu < 1$ or $\nu > 1$, the particle undergoes sub-diffusion or super-diffusion, respectively. The mean-square displacement for a general orientation-dependent motility is given by 
\begin{equation}%
\frac{\mean{ \Del{\vec{r}^{2}(t)} }}{ L^{2} } =  \frac{4 D_{\text{R}} t}{\text{Pe}^{2}}  + \sum_{\substack{k_{1}=-\infty \\ k_{1}\neq 0}}^{\infty} \sum_{\substack{k_{2}=-\infty \\ k_{2}\neq 0}}^{\infty} \vec{c}_{k_{1}} \cdot \vec{c}_{k_{2}} \left( \fct{C}_{k_{1}k_{2}} (D_{\text{R}} t) + \fct{C}_{k_{2}k_{1}} (D_{\text{R}} t) \right) \, e^{\ii (k_{1} + k_{2}) \phi_{0}}%
\label{eq_mean_squared_displacement}%
\end{equation}%
with
\begin{equation}%
\fct{C}_{k_{1} k_{2}}(t) = \begin{cases}
\frac{1}{k_{2}^{\,4}} \Big( k_{2}^{\,2} t - \big( 1 - e^{-k_{2}^{\,2} t} \big) \Big), & \text{for }k_{1} = - k_{2}, \\
\frac{1}{k_{2}^{\,4}} \Big( 1 - \big( 1 + k_{2}^{\,2} t \big) e^{-k_{2}^{\,2} t} \Big), & \text{for }k_{1} = - 2 k_{2}, \\
\frac{1}{ k_{1} (k_{1} + 2 k_{2}) } \Big( \frac{1}{k_{2}^{\,2}} \big( 1 - e^{-k_{1}^{\,2} t} \big) - \frac{1}{(k_{1} +k_{2})^2} \big( 1 - e^{-(k_{1} +k_{2})^2 t} \big) \Big), & \, \text{else.}
\end{cases}
\end{equation}
In Figs.~\ref{fig3}h and \ref{fig4}h, we compare the experimentally determined mean-square displacement with the corresponding theoretical result. We observe three temporal regimes, characterized by two crossover times. By expanding the analytic result for the mean-square displacement in time, we obtain 
\begin{equation}%
\mean{\Del{\vec{r}}^{2}(t)}  = 4 D_{ \text{T} } t +  \vec{v}^{2}(\phi_0)  t^2 + D_{ \text{R} } \Big( 3 \partial_{\phi_0}^{2} \vec{v}^{2}(\phi_0) - 2 \big( \partial_{\phi_0} \vec{v}(\phi_0) \big)^{2} \Big) \frac{t^3}{6} + \mathcal{O}(t^4),
\label{eq_mean_squared_displacement_short_times}%
\end{equation}
where $\partial_{\phi_{0}}$ denotes the partial derivative with respect to the initial orientation $\phi_{0}$. Thus, the mean-square displacement starts in a short-time diffusion regime ($\nu = 1$), increasing linearly in time with the short-time diffusion coefficient $D_{ \text{S} } =  D_{ \text{T} }$. A transition from the short-time diffusive regime into a super-diffusive regime ($\nu > 1$) occurs, if the deterministic swimming motion dominates translational diffusion. This condition is fulfilled for times $t$ greater than the translational diffusion time $\tau_{ \text{D} } = D_{ \text{T} } / \vec{v}^{2}(\phi_0) $. As shown in Fig.~\ref{fig3}h, the transition to an intermediate super-diffusive regime is sensitive with respect to the initial velocity. If the particle is oriented initially along directions of high motility (see Fig.~\ref{fig3}h for $\phi_{0} = \pi/2$), the mean-square displacement displays a crossover to the ballistic regime ($\nu = 2$). However, if the initial velocity of the particle is not large enough to dominate translational diffusion or even vanishes (see Eq.\ \eqref{eq_mean_squared_displacement_short_times}), we observe a delayed crossover (see Fig.~\ref{fig3}h for $\phi_{0} = 0$). In that case, the particle has to undergo an angular displacement first, such that its propulsion grows until it overcomes translational diffusion. Due to this multiplicative coupling of diffusive and ballistic behavior for the angular and positional displacements, respectively, the mean-square displacement shows a super-ballistic power-law behavior ($\nu = 3$), which is masked by finite translational diffusion (see Eq.\ \eqref{eq_mean_squared_displacement_short_times}). For times $t$ greater than the rotational diffusion time $\tau_{ \text{R} } = 1 / D_{ \text{R} }$, the mean-square displacement evolves toward the diffusive limit ($\nu = 1$) again and it is described by a long-time diffusion coefficient 
\begin{equation}%
D_{\text{L}} = \lim\limits_{t \to \infty} \frac{ \mean{ \Del{\vec{r}}^{2}(t) } }{4 t} = D_{\text{T}} +  \frac{\bar{v}^{\,2}}{D_{\text{R}}} \sum_{k_{1}=1}^{\infty} \frac{\abs{ \vec{c}_{k_{1}} }^{2}}{k_{1}^{\,2}}. %
\end{equation}%
In the two experimental scenarios, the long-time diffusion coefficients are $D_{ \text{L},1} = 5.1 \,\mu\text{m}^{2}/\text{s}$ and $D_{ \text{L},2} = 2.6 \,\mu\text{m}^{2}/\text{s}$, respectively.

\subsection{Non-Gaussian parameter}
Finally, we study the non-Gaussian features of our active dynamics in more detail. Hence, we introduce the non-Gaussian parameter, which is defined in two spatial dimensions as \cite{KurzthalerF2017}
\begin{equation}%
\alpha_{2}(t) = \frac{1}{2} \frac{ \mean{ \Del{ \vec{r} }^{4}(t) } }{ \mean{ \Del{ \vec{r} }^{2}(t) }^{2} } - 1. %
\end{equation}%
The non-Gaussian parameter quantifies how far the distribution of displacements deviates from a Gaussian, i.e., $\alpha_{2}(t) = 0$ for an isotropic Gaussian distribution. For $\alpha_{2}(t)<0$ or $\alpha_{2}(t)>0$, the underlying distribution has less or more pronounced tails, respectively. Interesting for active Brownian motion is the case of deterministic motion (no tails), for which the non-Gaussian parameter is $\alpha_{2}(t)= -1/2$. To derive the analytic expression for the non-Gaussian parameter from our theoretical model, in addition to the mean-square displacement $\mean{ \Del{\vec{r}}^{2}(t)}$ the mean-quartic displacement $\mean{ \Del{\vec{r}}^{4}(t)}$ is also required, which is explicitly calculated in the Appendix. In Figs.~\ref{fig3}h and \ref{fig4}h, the anisotropy of the non-Gaussian behavior is visualized. For very small times $t\ll\tau_{ \text{D} }$, the displacements are simply diffusive, i.e., Gaussian, thus the non-Gaussian parameter $\alpha_{2}(t)$ is zero. For intermediate times $\tau_{ \text{D} }<t<\tau_{ \text{R} }$, the non-Gaussian parameter behaves anisotropically with respect to the initial orientation $\phi_{0}$. For a sufficiently high initial velocity, $\alpha_{2}(t)$ turns negative, which is characteristic for persistently swimming Brownian particles (see Fig.\ \ref{fig3}h for $\phi_{0}=\pi/2$). When the initial velocity vanishes, i.e., $\vec{v}(\phi_{0})=0$ (see Fig.\ \ref{fig3}h for $\phi_{0}=0$), we observe a positive non-Gaussian parameter. In this case, the particle moves mostly diffusively even for intermediate times, except for rare events where a fluctuation rotates the particle sufficiently, such that it experiences a large ballistic step. The underlying distribution of displacements is thus Gaussian with pronounced tails which dominate the fourth moment over the second and lead to a positive non-Gaussianity. Finally, for long times $t>\tau_{ \text{R} }$, we observe a long-lived non-Gaussianity in the case of two-fold symmetry and Gaussian behavior in the case of four-fold symmetry. To explain this observation, we consider the covariance matrix of the displacement distribution and we define the long-time diffusion matrix
\begin{equation}%
\big(\vec{D}_{\text{L}}\big)_{ij} = \lim\limits_{t \to \infty} \frac{ \mean{ \Del{\fct{r}}_{i}(t)\Del{\fct{r}}_{j}(t) } }{2 t} = D_{\text{T}} \delta_{ij} +  \frac{\bar{v}^{\,2}}{D_{\text{R}}} \sum_{k_{1}=1}^{\infty} \frac{ 1 }{ k_{1}^{\,2}} \left( \fct{c}_{k_{1},i} \fct{c}_{-k_{1},j} + \fct{c}_{-k_{1},i} \fct{c}_{k_{1},j} \right) %
\end{equation}%
for $i,j \in \{x,y\}$. The eigenvalues of this matrix are given as $D_{\pm} = D_{ \text{L} } \pm \Del{D}_{ \text{L} }$, where $\Del{D}_{ \text{L} }$ denotes the long-time anisotropy
\begin{equation}%
\Del{D}_{ \text{L} } =\frac{\bar{v}^{\,2}}{D_{\text{R}}} \sqrt{ \sum_{k_{1}=1}^{\infty} \sum_{k_{2}=1}^{\infty} \frac{1}{k_{1}^{\,2} k_{2}^{\,2}} \left( \abs{ \vec{c}_{k_{1}} \cdot \vec{c}_{k_{2}} }^{2} + \abs{ \vec{c}_{k_{1}} \cdot \vec{c}_{-k_{2}} }^{2} - \abs{ \vec{c}_{k_{1}} }^{2} \abs{ \vec{c}_{k_{2}} }^{2} \right)  },%
\end{equation}%
which describes the long-time diffusion along the principal axes of maximal and minimal diffusion, respectively. In the two experimental scenarios, the long-time anisotropy yields $\Del{D}_{ \text{L},1} = 4.0\,\mu\text{m}^{2}/\text{s}$ and $\Del{D}_{ \text{L},2} = 0 \,\mu\text{m}^{2}/\text{s}$, respectively. Using the introduced notation, the long-time behavior of the non-Gaussian parameter can be expressed as
\begin{equation}%
\lim\limits_{t \to \infty}  \alpha_{2}(t) = \frac{1}{2} \left( \frac{ \Del{D}_{ \text{L} } }{ D_{ \text{L} } } \right)^{2}, %
\end{equation}%
which coincides with the non-Gaussianity of an anisotropic Gaussian distribution with covariance matrix $2 \vec{D}_{\text{L}} t$. Thus, the long-time behavior of the non-Gaussian parameter quantifies the anisotropy of the long-time diffusion. For the motility with two-fold symmetry, we have an enhanced long-time diffusion along the $y$-axis and a decreased long-time diffusion along the $x$-axis leading to a non-Gaussianity for long times (see Fig.~\ref{fig3}h). In the second scenario, the long-time behavior can be described with solely one long-time diffusion coefficient, thus the non-Gaussian parameter vanishes (see Fig.~\ref{fig4}h).

\pagebreak
\section{\label{V}Conclusions} 
In this work, we reported on a new methodology to impose a complex anisotropic motility behavior to active Brownian particles. We engineered an orientation-dependent motility of active dumbbells whose rotational diffusivity is externally controlled by randomized magnetic fields and whose propulsion velocity is prescribed using a feedback scheme, which updates the velocity based on the particles' orientation. To describe the dynamical features of the particles, we developed a theoretical framework, which proved to be in good agreement with corresponding experimental data. In particular, a particle's mean displacement shows deterministic active motion at very short times, decorrelation at intermediate times, and saturation to anisotropic persistence trajectories at long times. The mean-square displacement is also characterized by different temporal regimes. We found that the transition from isotropic diffusion at short times to a super-diffusive intermediate regime is very sensitive to the initial velocity of the particle such that the coupling of diffusive-rotational and ballistic-translational motion can result in super-ballistic motion. Moreover, the motion is characterized by anisotropic diffusion at long times, as described by the long-time diffusion coefficient and the long-time anisotropy. Finally, we have investigated the deviation from a standard Gaussian distribution by calculating the non-Gaussian parameter as a function of time. It becomes non-zero for intermediate times: negative when there is persistent swimming, and positive during reorientation events from an initial orientation with low velocity to orientations with high velocity. Furthermore, the long-time behavior quantifies the anisotropy of the long-time diffusion, being non-zero for the two-fold-symmetric motility and zero for the four-fold-symmetric motility.

The basic model we proposed here is applicable to a broad range of systems with anisotropic external propulsion mechanisms and relevant in the context of the orientational dependence of the propulsion speed, which can intrinsically emerge for both artificial and biological microswimmers \cite{Uspal2019,HuWWG2015}. In the future, intricate combinations of spatial, orientational, and temporal modulations of motility could be considered. One could also proceed to particles with a complex shape, which have more involved trajectories.\cite{WittkowskiL2012,VossW2018} Finally, although in our current experiments one particle at a time is controlled, we envision possible experimental realizations to control many particles to explore emerging collective effects \cite{CohenG2014}.

\pagebreak
\begin{acknowledgement}
R.W.\ and H.L.\ are funded by the Deutsche Forschungsgemeinschaft (DFG, German Research Foundation) -- WI 4170/3-1; LO 418/23-1. M.A.F.R., L.A., and L.I. acknowledge financial support from the Swiss National Science Foundation Grant PP00P2-172913/1.
\end{acknowledgement}

\pagebreak
\section{Appendix}

\subsection{Real-time feedback}
\begin{figure}[htb]
\centering
\includegraphics[width=\columnwidth]{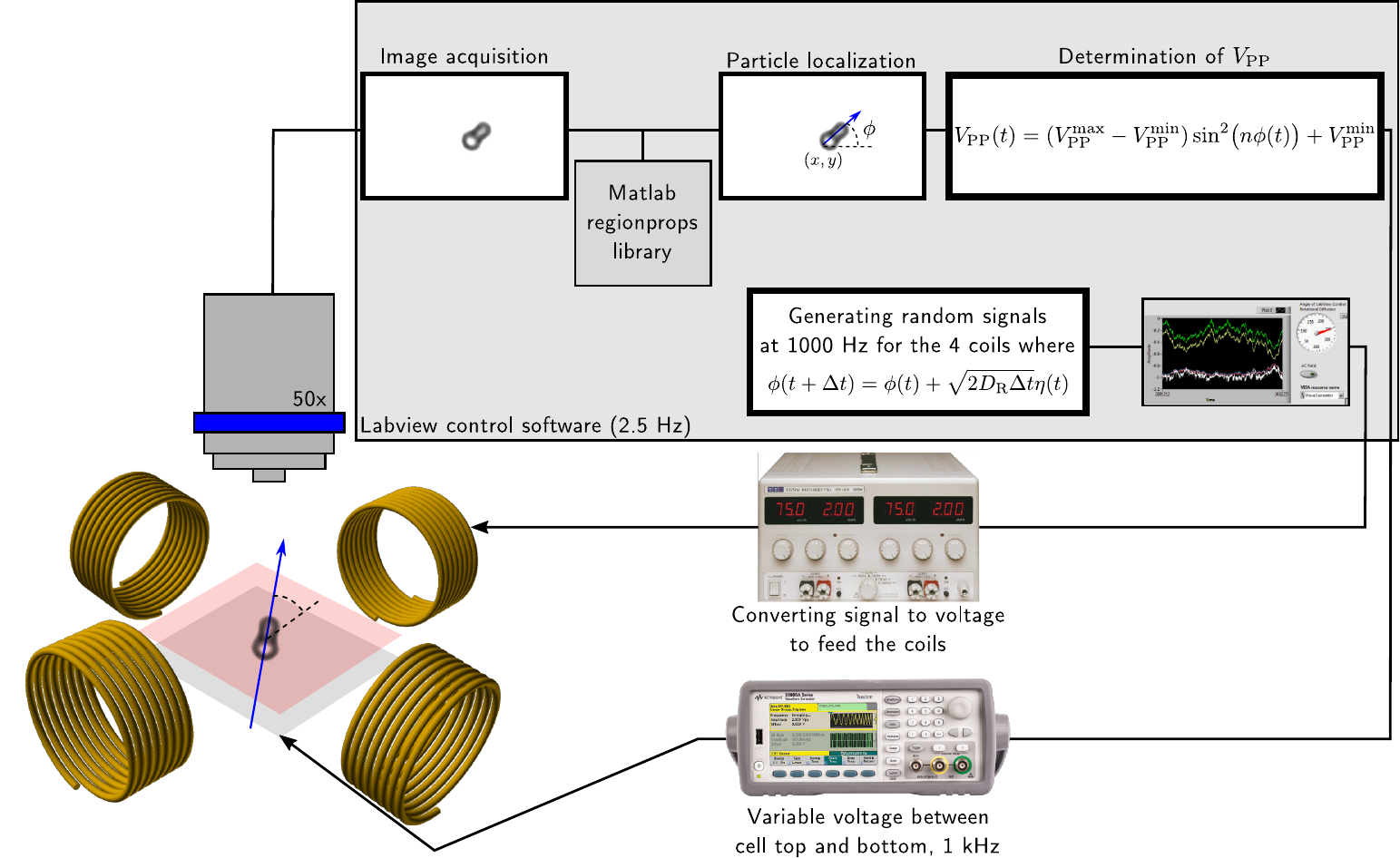}
\caption*{\label{figA1}Figure A1. Scheme of the real-time feedback applied in the experiments.}
\end{figure}

\subsection{Post-processing and data analysis} 
We collected 45 trajectories (86 min recording time in total) with a propulsion velocity with two-fold symmetry and 15 trajectories (28 min recording time in total) with a propulsion velocity with four-fold symmetry. The position $\vec{r}$ and the orientation $\phi$ were recorded at 2.5 fps and the velocity was calculated from the displacement of successive positions of the particle as $\vec{v}(t) = \left( \vec{r}(t +\Del{t}) - \vec{r}(t) \right)/\Del{t}$, where $\Del{t} = 0.4\,\text{s}$ is the time between two frames. The time steps are not fully equidistant, therefore the experimental data were linearly interpolated to obtain equidistant points. Initially, we did not distinguish each lobe of the dumbbell and thus we measured its orientation modulo $\pi$. From the direction of the velocity we could post-process the trajectory to reconstruct the angles in the interval [$0,2\pi$). Finally, we rescaled all displacements with a characteristic length $L = \bar{v}/D_{ \text{R} }$ and all times with the inverse rotational diffusion coefficient $1/D_{ \text{R} }$, where $\bar{v}$ is the orientationally averaged speed for a trajectory. Experimental means with respect to a specific initial orientation $\phi_{0}$ were calculated by averaging in the interval $[\phi_{0} - \del{\phi}, \phi_{0} + \del{\phi}]$. We chose $\del{\phi} = 25^{\circ}$ and modified the theoretical results accordingly by $\exp(\ii k \phi ) \to \exp(\ii k \phi ) \sin(k \del{\phi}) /(k \del{\phi})  $. In Figs.~\ref{fig3}g-h and \ref{fig4}g-h, we took advantage of the rotational and inflection symmetries of the experiment to increase the statistics.

\subsection{General theoretical result}
In this section, we calculate the $n$-th moment of the translational displacement $\mean{\Del{\vec{r}}^{n}(t)} = \mean{(\vec{r}(t) - \vec{r}_{0})^{n}}$ for active Brownian motion with a general orientation-dependent motility. With respect to initial conditions $\vec{r}(0)=\vec{r}_{0}$ and $\phi(0)=\phi_{0}$, solutions to the Langevin equations (\ref{eq_langevin_trans}) and \eqref{eq_langevin_rot} are obtained via simple integration as
{\allowdisplaybreaks
	\begin{align}%
	\vec{r}(t) & = \vec{r}_{0} + \integral{0}{t}{ \Big( \vec{v}\big( \phi(t') \big) + \sqrt{2 D_{\text{T}} } \vec{\xi}(t') \Big) }{t'}, \label{eq_langevin_trans_solution} \\%
	\phi(t) & = \phi_{0} + \sqrt{2 D_{\text{R}} } \integral{0}{t}{ \eta(t') }{ t' }.
	\label{eq_langevin_rot_solution}%
\end{align}}%
Since $\phi(t)$ is a linear combination of Gaussian variables, the corresponding probability distribution is Gaussian as well and the conditional probability density $\fct{P}(\phi_{2},t_{2}\vert\phi_{1},t_{1})$ is given by 
\begin{equation}%
\fct{P}(\phi_{2},t_{2}\vert\phi_{1},t_{1}) = \frac{1}{ \sqrt{ 4 \pi  D_{\text{R}} (t_{2}-t_{1}) } } \exp\left( - \frac{ ( \phi_{2} - \phi_{1} )^2 }{ 4  D_{\text{R}} (t_{2}-t_{1}) }\right).
\label{eq_distribution_rot}
\end{equation}%
The conditional probability density $\fct{P}(\phi_{2},t_{2}\vert\phi_{1},t_{1})$ embodies the probability of finding the particle with orientation $\phi_{2}$ at time $t_{2}$ under the condition that the particle was oriented at an angle $\phi_{1}$ at former time $t_{1}$. Next, we construct the joint probability density of finding the particle at an angle $\phi_{1}$ at time $t_{1}$, at an angle $\phi_{2}$ at time $t_{2}$, \dots, and at an angle $\phi_{n}$ at time $t_{n}$ as $\fct{P}(\phi_{n},t_{n}; \dots; \phi_{1},t_{1}) = \prod_{j=1}^{n} \fct{P}(\phi_{j},t_{j}\vert\phi_{j-1},t_{j-1})$ using the Markovian property of the Gaussian white noise. The knowledge of the joint probability density $\fct{P}(\phi_{n},t_{n}; \dots; \phi_{1},t_{1})$ allows for an analytic calculation of the $n$-th moment of the translational displacement $\mean{\Delta \vec{r}(t)^{n}}$. The translational displacement $ \Del{\vec{r}}(t) = \Del{\vec{r}}_{ \text{A} }(t) + \Del{\vec{r}}_{ \text{D} }(t) $ can be split into an active contribution $\Del{\vec{r}}_{ \text{A} }(t) = \integral{0}{t}{ \vec{v}\big( \phi(t_{1}) \big) }{t_{1}} $ and a diffusive contribution $ \Del{\vec{r}}_{ \text{D} }(t)  = \sqrt{2 D_{ \text{T} } } \integral{0}{t}{ \vec{\xi}(t_{1})  }{t_{1}}$. These two parts are stochastically independent and therefore the $n$-th moment of the total displacement can be represented as
{\allowdisplaybreaks
\begin{align}%
		& \mean{ \Del{\vec{r}}^{2n}(t) } =  \sum_{n_{1} + 2 n_{2} + n_{3} = n }   \frac{n!}{n_{1}! n_{2}! n_{2}! n_{3} !} \mean{ \Del{\vec{r}}_{ \text{A} }^{2(n_{1} + n_{2})}(t) }  \mean{ \Del{\vec{r}}_ \text{D}^{2(n_{2} + n_{3})}(t) },  \\ %
    	& \mean{ \Del{\vec{r}}^{2n+1}(t) } =  \sum_{n_{1} + 2 n_{2} + n_{3} = n }   \frac{n!}{n_{1}! n_{2}! n_{2}! n_{3} !} \mean{ \Del{\vec{r}}_{ \text{A} }^{2(n_{1} + n_{2})+1}(t) }  \mean{ \Del{\vec{r}}_{ \text{D} }^{2(n_{2} + n_{3})}(t) }  \\ %
    	& \qquad\qquad\qquad\; + \sum_{n_{1} + 2 n_{2} + n_{3} = n - 1}   \frac{n!}{n_{1}! n_{2}! (n_{2}+1)! n_{3} !} \mean{ \Del{\vec{r}}_{ \text{A} }^{2(n_{1} + n_{2})+1}(t) }  \mean{ \Del{\vec{r}}_{ \text{D} }^{2(n_{2} + n_{3} + 1)}(t) }. \nonumber%
\end{align}}%
Like the orientation angle, also the diffusive displacement $\Del{\vec{r}}_{\text{D}}(t)$ is a sum of Gaussian variables and hence it follows a Gaussian distribution. The corresponding moments are calculated as 
{\allowdisplaybreaks
\begin{align}%
	& \mean{ \Del{\vec{r}}_{ \text{D} }^{2n}(t) } = n! \left( 4 D_{\text{T}} t \right)^{n}, \\%
	& \mean{ \Del{\vec{r}}_{ \text{D} }^{2n + 1 }(t) } = \vec{0}. %
\end{align}}%
In contrast to the diffusive displacement, the active displacement $\Del{\vec{r}}_{ \text{A} }(t)$ is a nonlinear combination of Gaussian variables. Here, the joint probability density $\fct{P}(\phi_{n},t_{n}; \dots; \phi_{1},t_{1})$ is used to calculate the $n$-th moment as
{\allowdisplaybreaks
	\begin{equation}%
	\mean{ \Del{\vec{r}}_\text{A}^{n}(t) } =  L^n \sum_{\substack{k_{1}=-\infty \\ 
	k_{1}\neq 0}}^{\infty} \dotsi \sum_{\substack{k_{n}=-\infty \\ 
	k_{n}\neq 0}}^{\infty}  \vec{c}_{k_{1}} \cdot \, \dotsb \, \cdot \vec{c}_{k_{n}} e^{\ii \left( \sum_{j=1}^{n} k_{j} \right) \phi_{0}}  \sum_{\sigma \in S_{n}} \fct{C}_{k_{\sigma(1)} \dotsb k_{\sigma(n)}}(D_{\text{R}} t),%
	\end{equation}}%
where the sum has to be performed over the $n!$ permutations of the symmetric group $ S_{n}$ and
{\allowdisplaybreaks
	\begin{equation}%
	\fct{C}_{k_{1} \dotsb k_{n}}(t) = \integralvar{0}{t}{t_{n}} \integralvar{0}{t_{n}}{t_{n-1}}  \,\dotsi \integralvar{0}{t_{2}}{t_{1}} \prod_{j=1}^{n} e^{- \left( \sum_{l=j}^{n} k_{l} \right)^{2}  ( t_{j} - t_{j-1}) }. %
\end{equation}}%

\subsection{Low-order moments}
The low-order moments for Brownian motion with an orientation-dependent motility are 
\begin{align}%
&\frac{\mean{ \Del{\vec{r}(t) } } }{ L }  = \frac{\mean{ \Del{\vec{r}_{ \text{A} }(t)} }}{ L }, \\%
&\frac{\mean{ \Del{\vec{r}}^{2}(t) } }{ L^2 }  = \frac{4 D_{\text{R}} t}{\text{Pe}^{2}}  +  \frac{\mean{ \Del{\vec{r}}_\text{A}^{2}(t) }}{ L^2 },  \\ %
&\frac{\mean{ \Del{\vec{r}}^{4}(t) }}{ L^4 }  = \frac{32 (D_{\text{R}} t)^{2}}{\text{Pe}^{4}} + \frac{16 D_{\text{R}} t}{\text{Pe}^{2}}  \frac{\mean{ \Del{\vec{r}_{ \text{A} }^{2}(t)} }}{ L^2 } +  \frac{\mean{ \Del{\vec{r}}_{ \text{A} }^{4}(t) }}{ L^4 }. %
\end{align}%
For a propulsion velocity with two-fold symmetry $\vec{v}_{1}(\phi) =  2 \bar{v}_{1} \sin^{2}\left( \phi \right) \uvec{u}(\phi)$, with non-zero Fourier-coefficient vectors $\vec{c}_{-3} = -(1,\ii)/4$, $\vec{c}_{-1} = (1,3\ii)/4$, $\vec{c}_{1} = (1,-3\ii)/4$, and $\vec{c}_{3} = (-1,\ii)/4$, one obtains
{\allowdisplaybreaks
	\begin{align}%
	&\frac{\mean{ \Del{\vec{r}_{ \text{A} }(t)} }}{ L } \,\,  = \frac{1}{2} \left(1-e^{-\tau}\right) \begin{pmatrix} \cos ( \phi_{0} ) \\ 3 \sin ( \phi_{0} )  \end{pmatrix} - \frac{1}{18} \left(1-e^{-9 \tau}\right) \begin{pmatrix} \cos ( 3 \phi_{0} ) \\ \sin ( 3 \phi_{0} )  \end{pmatrix}, \\%
	&\frac{\mean{  \Del{\vec{r}}_{ \text{A} }^{2}(t)  }}{ L^{2} } = \frac{1}{162} \left( 414 \tau - 406 +405 e^{-\tau} +e^{-9 \tau}\right)  - \frac{\cos ( 2 \phi_{0} ) }{45} \left(35 - 45 e^{-\tau} + 9 e^{-4 \tau} + e^{-9 \tau} \right)  \nonumber \\%
	& \hspace{2.2cm} + \frac{\cos ( 4 \phi_{0} ) }{5040}  \left(175 - 168 e^{-\tau} -40 e^{-9 \tau} + 33 e^{-16 \tau} \right), \\%
	& \frac{\mean{ \Del{\vec{r}}_{ \text{A} }^{4}(t) }}{ L^4 } = \frac{1}{185177664} \Big( 6615 \left( 477619.2 \tau^2 - 1646424 \tau + 2289911.3 \right) - 2057529.6 (2025 \tau \nonumber  \\
	& \hspace{1.2cm} + 7376) e^{-\tau} + 27905245.2 e^{-4 \tau} - 6.4 (84105 \tau-92242) e^{-9 \tau} + 79388.1 e^{-16 \tau} + 98 e^{-36 \tau} \Big) \nonumber \\
	& - \frac{ \cos (2 \phi_{0}) }{28291032} \Big( 14553 (21396 \tau - 41812.87) + 314344.8 (854 \tau + 1986.9) e^{-\tau} - 80.19 (221648 \tau  \nonumber \\
	& + 203425.3) e^{-4 \tau} - 8.8 (37724.4 \tau - 24587.89) e^{-9 \tau} + 25776.63 e^{-16 \tau}  + 1422.36 e^{-25 \tau} + 116.375 e^{-36 \tau} \Big)  \nonumber \\
	& + \frac{\cos (4\phi_{0}) }{1287241956} \Big( 27027 (30135 \tau - 50390.53) + 119189.07 (4296 \tau + 1143.67) e^{-\tau} + 4580265.69 e^{-4 \tau}  \nonumber \\
	& + 7.15 (999432 \tau - 779845.1) e^{-9 \tau} - 21.06 (128898 \tau + 27634.9) e^{-16 \tau}  + 340798.185 e^{-25 \tau} \nonumber \\ 
	&  \hspace{2.9cm} + 10718.75 e^{-36 \tau}   + 1318.275 e^{-49 \tau} \Big) \nonumber \\
	& -\frac{ \cos (6 \phi_{0})}{3895779888} \Big( 122694 (196 \tau - 70.67) + 1082161.08 e^{-\tau} + 8541342.81 e^{-4 \tau} + 8179.6 (112 \tau \nonumber \\
	& - 130.1) e^{-9 \tau} + 563165.46 e^{-16 \tau} - 565994.52 e^{-25 \tau} - 49 (1144 \tau - 1918.79) e^{-36 \tau} + 20255.4 e^{-49 \tau} \Big) \nonumber \\
	& + \frac{\cos (8 \phi_{0}) }{217945728} \Big( 24277.11 - 31231.2 e^{-\tau} + 7207.2 e^{-4 \tau} -2955.68 e^{-9 \tau} + 4204.2 e^{-16 \tau}  \nonumber \\
	& \hspace{2.7cm} - 1515.36 e^{-25 \tau} + 308 e^{-36 \tau} - 491.04 e^{-49 \tau} + 196.77 e^{-64 \tau} \Big)
	\end{align}}%
with $L = \bar{v}_{1}/ D_{\text{R}}$ and  $\tau = D_{\text{R}} t$. In the case of a propulsion velocity with four-fold symmetry $\vec{v}_{2}(\phi) =  2 \bar{v}_{2} \sin^{2}\left(2 \phi \right) \uvec{u}(\phi)$, with non-zero Fourier-coefficient vectors $\vec{c}_{-5} = -(1,\ii)/4$, $\vec{c}_{-3} = (-1,\ii)/4$, $\vec{c}_{-1} = (-1,\ii)/2$, $\vec{c}_{1} = (1,-\ii)/2$, $\vec{c}_{3} = - (1,\ii)/4$, and $\vec{c}_{5} = (-1,\ii)/4$, one finds
{\allowdisplaybreaks
	\begin{align}%
	&\frac{\mean{ \Del{\vec{r}_{ \text{A} }(t)} }}{ L }  =  \left(1-e^{-\tau}\right) \begin{pmatrix} \cos ( \phi_{0} ) \\  \sin ( \phi_{0} )  \end{pmatrix} - \frac{1}{18} \left(1-e^{-9 \tau}\right) \begin{pmatrix} \cos ( 3 \phi_{0} ) \\ - \sin ( 3 \phi_{0} )  \end{pmatrix} - \frac{1}{50} \left(1-e^{-25 \tau}\right) \begin{pmatrix} \cos ( 5 \phi_{0} ) \\ \sin ( 5 \phi_{0} )  \end{pmatrix},\\%
	& \frac{\mean{  \Del{\vec{r}}_{ \text{A} }^{2}(t)  }}{ L^{2} } =  \frac{1}{101250} \left(210150 \tau - 203206 + 202500 e^{-\tau} + 625 e^{-9 \tau} + 81 e^{-25 \tau} \right) \nonumber \\%
	& \hspace{2.2cm} - \frac{\cos (4\phi_{0})}{6300 } \left(847 -840 e^{-\tau}  - 100 e^{-9 \tau} + 65 e^{-16 \tau} + 28 e^{-25 \tau} \right)  \nonumber \\%
	& \hspace{2.2cm} + \frac{ \cos (8 \phi_{0})}{2059200} \left( 2431 - 2080 e^{-9 \tau} - 1056 e^{-25 \tau} + 705 e^{-64 \tau} \right), \\%
	& \frac{\mean{ \Del{\vec{r}}_{ \text{A} }^{4}(t) }}{ L^4 } = \frac{1}{473337256392}  \Big( 2705.4027 \big( 1507431168 \tau^2 - 5233851943.2 \tau + 7368029142.0341 \big) \nonumber \\
	&  - 858661689.6 ( 6657 \tau + 23236 ) e^{-\tau} + 18488059504.2 e^{-4 \tau} - 135.2 (12871089 \tau + 1792537.7) e^{-9 \tau}  \nonumber \\
	& + 100773489.96 e^{-16 \tau} - 3.73248 (17191424.25 \tau  - 5307384.493) e^{-25 \tau} + 10678232.89 e^{-36 \tau} \nonumber \\ 
	& \hspace{5.1cm} + 110963.93765625 e^{-64 \tau} + 4207.44227904 e^{-100 \tau} \Big) \nonumber \\
	& - \frac{ \cos (4 \phi_{0})}{29728209352842} \Big( 20805.83505 (1594853568 \tau - 2813134015.1) + 898812074.16 (26628 \tau \nonumber \\
	& +65452.9) e^{-\tau} - 2689320626.8929 e^{-4 \tau} + 10917.4 ( 25742178 \tau + 588955.9) e^{-9 \tau} \nonumber \\
	& - 302.328 (353017665 \tau + 138167563) e^{-16 \tau} - 65.1168 (343828485 \tau - 31369636.51 ) e^{-25 \tau} \nonumber \\
	& + 1905187479.104 e^{-36 \tau} +32724549.585 e^{-49 \tau} + 21910524.654375 e^{-64 \tau} + 13975832.871 e^{-81 \tau} \nonumber \\
	& \hspace{3.8cm} + 3112728.130512 e^{-100 \tau} \Big) \nonumber \\
	& + \frac{\cos (8 \phi_{0})}{2255833858668831441} \Big( 154412.13645 (143187946902 \tau - 107200347827.9) \nonumber \\
	& + 14390958251825349.48 e^{-\tau} + 1820467586121508.467 e^{-4 \tau} + 6589810.87375 (205937424 \tau  \nonumber \\
	& + 56996291.3) e^{-9 \tau} + 13520156814211.905 e^{-16 \tau} + 712671.399 (275062788 \tau -12848345.233) e^{-25 \tau} \nonumber \\
	&  - 25027863332221.3925 e^{-36 \tau} - 10926088510221.027 e^{-49 \tau} - 2.32875 (15961369498074 \tau  \nonumber \\ 
	& + 4924072925496.7) e^{-64 \tau} + 8109806453743.995 e^{-81 \tau} + 874623947689.664172 e^{-100 \tau}  \nonumber \\
	&  \hspace{4.7cm} + 72681246322.059375 e^{-121 \tau} + 14384547897.773625 e^{-169 \tau} \Big) \nonumber \\
	& - \frac{\cos (12 \phi_{0})}{781427217274704} \Big( 57715889002.165 - 64186495813.44 e^{-\tau} + 4134535541.136 e^{-4 \tau} \nonumber \\
	& - 1797741837.892 e^{-9 \tau} + 2877730703.3475 e^{-16 \tau} + 706117054.5 e^{-25 \tau} + 1193818065.44 e^{-36 \tau} \nonumber \\
	&  - 298802203.128 e^{-49 \tau} + 128580438.987 e^{-64 \tau} - 379027122.76 e^{-81 \tau} - 90311261.04 e^{-100 \tau} \nonumber \\
	& \hspace{4.cm} - 124790292.9 e^{-121 \tau} + 78641802.3677 e^{-144 \tau} + 41855923.2168 e^{-169 \tau} \Big) \nonumber \\
	& +\frac{\cos (16 \phi_{0})}{3836848982174208} \Big(459813765.7869 -523542493.76 e^{-9 \tau} -178637472.4992 e^{-25 \tau} \nonumber \\ 
	& + 105544799.36 e^{-36 \tau}   + 185188888.17 e^{-64 \tau} +23763464.5248 e^{-100 \tau} -52195257.92 e^{-121 \tau} \nonumber \\
	& \hspace{4.2cm} -29527988.16 e^{-169 \tau}  + 9592294.4975 e^{-256 \tau} \Big) 
	\end{align}}%
with $L = \bar{v}_{2}/ D_{\text{R}}$ and  $\tau = D_{\text{R}} t$.

\begin{suppinfo}

\section{Supplementary Movie 1}
Recording of a representative trajectory of a dumbbell with a motility with two-fold symmetry. 

\section{Supplementary Movie 2}
Recording of a representative trajectory of a dumbbell with a motility with four-fold symmetry. 

\end{suppinfo}

\bibliography{refs}

\providecommand{\latin}[1]{#1}
\makeatletter
\providecommand{\doi}
  {\begingroup\let\do\@makeother\dospecials
  \catcode`\{=1 \catcode`\}=2 \doi@aux}
\providecommand{\doi@aux}[1]{\endgroup\texttt{#1}}
\makeatother
\providecommand*\mcitethebibliography{\thebibliography}
\csname @ifundefined\endcsname{endmcitethebibliography}
  {\let\endmcitethebibliography\endthebibliography}{}
\begin{mcitethebibliography}{63}
\providecommand*\natexlab[1]{#1}
\providecommand*\mciteSetBstSublistMode[1]{}
\providecommand*\mciteSetBstMaxWidthForm[2]{}
\providecommand*\mciteBstWouldAddEndPuncttrue
  {\def\EndOfBibitem{\unskip.}}
\providecommand*\mciteBstWouldAddEndPunctfalse
  {\let\EndOfBibitem\relax}
\providecommand*\mciteSetBstMidEndSepPunct[3]{}
\providecommand*\mciteSetBstSublistLabelBeginEnd[3]{}
\providecommand*\EndOfBibitem{}
\mciteSetBstSublistMode{f}
\mciteSetBstMaxWidthForm{subitem}{(\alph{mcitesubitemcount})}
\mciteSetBstSublistLabelBeginEnd
  {\mcitemaxwidthsubitemform\space}
  {\relax}
  {\relax}

\bibitem[Bechinger \latin{et~al.}(2016)Bechinger, {Di Leonardo}, L{\"o}wen,
  Reichhardt, Volpe, and Volpe]{BechingerdLLRVV2016}
Bechinger,~C.; {Di Leonardo},~R.; L{\"o}wen,~H.; Reichhardt,~C.; Volpe,~G.;
  Volpe,~G. Active particles in complex and crowded environments. \emph{Reviews
  of Modern Physics} \textbf{2016}, \emph{88}, 045006\relax
\mciteBstWouldAddEndPuncttrue
\mciteSetBstMidEndSepPunct{\mcitedefaultmidpunct}
{\mcitedefaultendpunct}{\mcitedefaultseppunct}\relax
\EndOfBibitem
\bibitem[Ebbens and Howse(2010)Ebbens, and Howse]{Howse2010}
Ebbens,~S.~J.; Howse,~J.~R. In pursuit of propulsion at the nanoscale.
  \emph{Soft Matter} \textbf{2010}, \emph{6}, 726--738\relax
\mciteBstWouldAddEndPuncttrue
\mciteSetBstMidEndSepPunct{\mcitedefaultmidpunct}
{\mcitedefaultendpunct}{\mcitedefaultseppunct}\relax
\EndOfBibitem
\bibitem[{Romanczuk} \latin{et~al.}(2012){Romanczuk}, {B{\"a}r}, {Ebeling},
  {Lindner}, and {Schimansky-Geier}]{RomanczukBELSG2012}
{Romanczuk},~P.; {B{\"a}r},~M.; {Ebeling},~W.; {Lindner},~B.;
  {Schimansky-Geier},~L. Active {B}rownian particles. {F}rom individual to
  collective stochastic dynamics. \emph{European Physical Journal Special
  Topics} \textbf{2012}, \emph{202}, 1--162\relax
\mciteBstWouldAddEndPuncttrue
\mciteSetBstMidEndSepPunct{\mcitedefaultmidpunct}
{\mcitedefaultendpunct}{\mcitedefaultseppunct}\relax
\EndOfBibitem
\bibitem[{Elgeti} \latin{et~al.}(2015){Elgeti}, {Winkler}, and
  {Gompper}]{ElgetiWG2015}
{Elgeti},~J.; {Winkler},~R.~G.; {Gompper},~G. Physics of microswimmers--single
  particle motion and collective behavior: a review. \emph{Reports on Progress
  in Physics} \textbf{2015}, \emph{78}, 056601\relax
\mciteBstWouldAddEndPuncttrue
\mciteSetBstMidEndSepPunct{\mcitedefaultmidpunct}
{\mcitedefaultendpunct}{\mcitedefaultseppunct}\relax
\EndOfBibitem
\bibitem[{Z{\"o}ttl} and {Stark}(2016){Z{\"o}ttl}, and {Stark}]{ZoettlS2016}
{Z{\"o}ttl},~A.; {Stark},~H. Emergent behavior in active colloids.
  \emph{Journal of Chemical Physics} \textbf{2016}, \emph{28}, 253001\relax
\mciteBstWouldAddEndPuncttrue
\mciteSetBstMidEndSepPunct{\mcitedefaultmidpunct}
{\mcitedefaultendpunct}{\mcitedefaultseppunct}\relax
\EndOfBibitem
\bibitem[Paxton \latin{et~al.}(2004)Paxton, Kistler, Olmeda, Sen, St.~Angelo,
  Cao, Mallouk, Lammert, and Crespi]{PaxtonKOSSCMLC2004}
Paxton,~W.~F.; Kistler,~K.~C.; Olmeda,~C.~C.; Sen,~A.; St.~Angelo,~S.~K.;
  Cao,~Y.; Mallouk,~T.~E.; Lammert,~P.~E.; Crespi,~V.~H. Catalytic nanomotors:
  autonomous movement of striped nanorods. \emph{Journal of the American
  Chemical Society} \textbf{2004}, \emph{126}, 13424--13431\relax
\mciteBstWouldAddEndPuncttrue
\mciteSetBstMidEndSepPunct{\mcitedefaultmidpunct}
{\mcitedefaultendpunct}{\mcitedefaultseppunct}\relax
\EndOfBibitem
\bibitem[Palacci \latin{et~al.}(2010)Palacci, Cottin-Bizonne, Ybert, and
  Bocquet]{PalacciCYB2010}
Palacci,~J.; Cottin-Bizonne,~C.; Ybert,~C.; Bocquet,~L. Sedimentation and
  effective temperature of active colloidal suspensions. \emph{Physical Review
  Letters} \textbf{2010}, \emph{105}, 088304\relax
\mciteBstWouldAddEndPuncttrue
\mciteSetBstMidEndSepPunct{\mcitedefaultmidpunct}
{\mcitedefaultendpunct}{\mcitedefaultseppunct}\relax
\EndOfBibitem
\bibitem[{Dietrich} \latin{et~al.}(2017){Dietrich}, {Renggli}, {Zanini},
  {Volpe}, {Buttinoni}, and {Isa}]{DietrichRZVBI2017}
{Dietrich},~K.; {Renggli},~D.; {Zanini},~M.; {Volpe},~G.; {Buttinoni},~I.;
  {Isa},~L. {Two-dimensional nature of the active Brownian motion of catalytic
  microswimmers at solid and liquid interfaces}. \emph{New Journal of Physics}
  \textbf{2017}, \emph{19}, 065008\relax
\mciteBstWouldAddEndPuncttrue
\mciteSetBstMidEndSepPunct{\mcitedefaultmidpunct}
{\mcitedefaultendpunct}{\mcitedefaultseppunct}\relax
\EndOfBibitem
\bibitem[Volpe \latin{et~al.}(2011)Volpe, Buttinoni, Vogt, K{\"u}mmerer, and
  Bechinger]{VolpeBVKB2011}
Volpe,~G.; Buttinoni,~I.; Vogt,~D.; K{\"u}mmerer,~H.-J.; Bechinger,~C.
  Microswimmers in patterned environments. \emph{Soft Matter} \textbf{2011},
  \emph{7}, 8810--8815\relax
\mciteBstWouldAddEndPuncttrue
\mciteSetBstMidEndSepPunct{\mcitedefaultmidpunct}
{\mcitedefaultendpunct}{\mcitedefaultseppunct}\relax
\EndOfBibitem
\bibitem[Buttinoni \latin{et~al.}(2012)Buttinoni, Volpe, K{\"u}mmel, Volpe, and
  Bechinger]{ButtinoniVKVB2012}
Buttinoni,~I.; Volpe,~G.; K{\"u}mmel,~F.; Volpe,~G.; Bechinger,~C. Active
  Brownian motion tunable by light. \emph{Journal of Physics: Condensed Matter}
  \textbf{2012}, \emph{24}, 284129\relax
\mciteBstWouldAddEndPuncttrue
\mciteSetBstMidEndSepPunct{\mcitedefaultmidpunct}
{\mcitedefaultendpunct}{\mcitedefaultseppunct}\relax
\EndOfBibitem
\bibitem[Palacci \latin{et~al.}(2013)Palacci, Sacanna, Steinberg, Pine, and
  Chaikin]{PalacciSSPC2013}
Palacci,~J.; Sacanna,~S.; Steinberg,~A.; Pine,~D.; Chaikin,~P. {Living crystals
  of light-activated colloidal surfers}. \emph{Science} \textbf{2013},
  \emph{339}, 936--940\relax
\mciteBstWouldAddEndPuncttrue
\mciteSetBstMidEndSepPunct{\mcitedefaultmidpunct}
{\mcitedefaultendpunct}{\mcitedefaultseppunct}\relax
\EndOfBibitem
\bibitem[Palacci \latin{et~al.}(2013)Palacci, Sacanna, Vatchinsky, Chaikin, and
  Pine]{PalacciSVCP2013}
Palacci,~J.; Sacanna,~S.; Vatchinsky,~A.; Chaikin,~P.~M.; Pine,~D.~J.
  Photoactivated colloidal dockers for cargo transportation. \emph{Journal of
  the American Chemical Society} \textbf{2013}, \emph{135}, 15978--15981\relax
\mciteBstWouldAddEndPuncttrue
\mciteSetBstMidEndSepPunct{\mcitedefaultmidpunct}
{\mcitedefaultendpunct}{\mcitedefaultseppunct}\relax
\EndOfBibitem
\bibitem[Palacci \latin{et~al.}(2014)Palacci, Sacanna, Kim, Yi, Pine, and
  Chaikin]{PalacciSKYPC2014}
Palacci,~J.; Sacanna,~S.; Kim,~S.-H.; Yi,~G.-R.; Pine,~D.~J.; Chaikin,~P.~M.
  Light-activated self-propelled colloids. \emph{Philosophical Transactions of
  the Royal Society A: Mathematical, Physical and Engineering Sciences}
  \textbf{2014}, \emph{372}, 20130372\relax
\mciteBstWouldAddEndPuncttrue
\mciteSetBstMidEndSepPunct{\mcitedefaultmidpunct}
{\mcitedefaultendpunct}{\mcitedefaultseppunct}\relax
\EndOfBibitem
\bibitem[Moyses \latin{et~al.}(2016)Moyses, Palacci, Sacanna, and
  Grier]{MoysesPSG2016}
Moyses,~H.; Palacci,~J.; Sacanna,~S.; Grier,~D.~G. Trochoidal trajectories of
  self-propelled Janus particles in a diverging laser beam. \emph{Soft Matter}
  \textbf{2016}, \emph{12}, 6357--6364\relax
\mciteBstWouldAddEndPuncttrue
\mciteSetBstMidEndSepPunct{\mcitedefaultmidpunct}
{\mcitedefaultendpunct}{\mcitedefaultseppunct}\relax
\EndOfBibitem
\bibitem[Wang \latin{et~al.}(2012)Wang, Castro, Hoyos, and
  Mallouk]{WangCHM2012}
Wang,~W.; Castro,~L.~A.; Hoyos,~M.; Mallouk,~T.~E. Autonomous motion of
  metallic microrods propelled by ultrasound. \emph{ACS Nano} \textbf{2012},
  \emph{6}, 6122--6132\relax
\mciteBstWouldAddEndPuncttrue
\mciteSetBstMidEndSepPunct{\mcitedefaultmidpunct}
{\mcitedefaultendpunct}{\mcitedefaultseppunct}\relax
\EndOfBibitem
\bibitem[Dreyfus \latin{et~al.}(2005)Dreyfus, Baudry, Roper, Fermigier, Stone,
  and Bibette]{DreyfusBRFSB2005}
Dreyfus,~R.; Baudry,~J.; Roper,~M.~L.; Fermigier,~M.; Stone,~H.~A.; Bibette,~J.
  Microscopic artificial swimmers. \emph{Nature} \textbf{2005}, \emph{437},
  862--865\relax
\mciteBstWouldAddEndPuncttrue
\mciteSetBstMidEndSepPunct{\mcitedefaultmidpunct}
{\mcitedefaultendpunct}{\mcitedefaultseppunct}\relax
\EndOfBibitem
\bibitem[Grosjean \latin{et~al.}(2015)Grosjean, Lagubeau, Darras, Hubert,
  Lumay, and Vandewalle]{GrosjeanLDHLV2015}
Grosjean,~G.; Lagubeau,~G.; Darras,~A.; Hubert,~M.; Lumay,~G.; Vandewalle,~N.
  Remote control of self-assembled microswimmers. \emph{Scientific Reports}
  \textbf{2015}, \emph{5}, 16035\relax
\mciteBstWouldAddEndPuncttrue
\mciteSetBstMidEndSepPunct{\mcitedefaultmidpunct}
{\mcitedefaultendpunct}{\mcitedefaultseppunct}\relax
\EndOfBibitem
\bibitem[Steinbach \latin{et~al.}(2016)Steinbach, Gemming, and
  Erbe]{SteinbachGE2016}
Steinbach,~G.; Gemming,~S.; Erbe,~A. Non-equilibrium dynamics of magnetically
  anisotropic particles under oscillating fields. \emph{European Physical
  Journal E} \textbf{2016}, \emph{39}, 69\relax
\mciteBstWouldAddEndPuncttrue
\mciteSetBstMidEndSepPunct{\mcitedefaultmidpunct}
{\mcitedefaultendpunct}{\mcitedefaultseppunct}\relax
\EndOfBibitem
\bibitem[Kaiser \latin{et~al.}(2017)Kaiser, Snezhko, and Aranson]{KaiserSA2017}
Kaiser,~A.; Snezhko,~A.; Aranson,~I.~S. Flocking ferromagnetic colloids.
  \emph{Science Advances} \textbf{2017}, \emph{3}, e1601469\relax
\mciteBstWouldAddEndPuncttrue
\mciteSetBstMidEndSepPunct{\mcitedefaultmidpunct}
{\mcitedefaultendpunct}{\mcitedefaultseppunct}\relax
\EndOfBibitem
\bibitem[Bricard \latin{et~al.}(2013)Bricard, Caussin, Desreumaux, Dauchot, and
  Bartolo]{BricardCDDB2013}
Bricard,~A.; Caussin,~J.~B.; Desreumaux,~N.; Dauchot,~O.; Bartolo,~D. Emergence
  of macroscopic directed motion in populations of motile colloids.
  \emph{Nature} \textbf{2013}, \emph{503}, 95--98\relax
\mciteBstWouldAddEndPuncttrue
\mciteSetBstMidEndSepPunct{\mcitedefaultmidpunct}
{\mcitedefaultendpunct}{\mcitedefaultseppunct}\relax
\EndOfBibitem
\bibitem[Morin \latin{et~al.}(2017)Morin, Desreumaux, Caussin, and
  Bartolo]{MorinDCB2017}
Morin,~A.; Desreumaux,~N.; Caussin,~J.-B.; Bartolo,~D. Distortion and
  destruction of colloidal flocks in disordered environments. \emph{Nature
  Physics} \textbf{2017}, \emph{13}, 63--67\relax
\mciteBstWouldAddEndPuncttrue
\mciteSetBstMidEndSepPunct{\mcitedefaultmidpunct}
{\mcitedefaultendpunct}{\mcitedefaultseppunct}\relax
\EndOfBibitem
\bibitem[K{\"u}mmel \latin{et~al.}(2013)K{\"u}mmel, {ten Hagen}, Wittkowski,
  Buttinoni, Eichhorn, Volpe, L{\"o}wen, and Bechinger]{KuemmeltHWBEVLB2013}
K{\"u}mmel,~F.; {ten Hagen},~B.; Wittkowski,~R.; Buttinoni,~I.; Eichhorn,~R.;
  Volpe,~G.; L{\"o}wen,~H.; Bechinger,~C. Circular motion of asymmetric
  self-propelling particles. \emph{Physical Review Letters} \textbf{2013},
  \emph{110}, 198302\relax
\mciteBstWouldAddEndPuncttrue
\mciteSetBstMidEndSepPunct{\mcitedefaultmidpunct}
{\mcitedefaultendpunct}{\mcitedefaultseppunct}\relax
\EndOfBibitem
\bibitem[K{\"u}mmel \latin{et~al.}(2014)K{\"u}mmel, {ten Hagen}, Wittkowski,
  Takagi, Buttinoni, Eichhorn, Volpe, L{\"o}wen, and
  Bechinger]{KuemmeltHWTBEVLB2014}
K{\"u}mmel,~F.; {ten Hagen},~B.; Wittkowski,~R.; Takagi,~D.; Buttinoni,~I.;
  Eichhorn,~R.; Volpe,~G.; L{\"o}wen,~H.; Bechinger,~C. Reply to ``{C}omment on
  `{C}ircular motion of asymmetric self-propelling particles'". \emph{Physical
  Review Letters} \textbf{2014}, \emph{113}, 029802\relax
\mciteBstWouldAddEndPuncttrue
\mciteSetBstMidEndSepPunct{\mcitedefaultmidpunct}
{\mcitedefaultendpunct}{\mcitedefaultseppunct}\relax
\EndOfBibitem
\bibitem[{Zheng} \latin{et~al.}(2013){Zheng}, {ten Hagen}, {Kaiser}, {Wu},
  {Cui}, {Silber-Li}, and {L{\"o}wen}]{ZhengtHKWCSLL2013}
{Zheng},~X.; {ten Hagen},~B.; {Kaiser},~A.; {Wu},~M.; {Cui},~H.;
  {Silber-Li},~Z.; {L{\"o}wen},~H. Non-{G}aussian statistics for the motion of
  self-propelled {J}anus particles: experiment versus theory. \emph{Physical
  Review E} \textbf{2013}, \emph{88}, 032304\relax
\mciteBstWouldAddEndPuncttrue
\mciteSetBstMidEndSepPunct{\mcitedefaultmidpunct}
{\mcitedefaultendpunct}{\mcitedefaultseppunct}\relax
\EndOfBibitem
\bibitem[Hong \latin{et~al.}(2007)Hong, Blackman, Kopp, Sen, and
  Velegol]{HongBKSV2007}
Hong,~Y.; Blackman,~N.~M.; Kopp,~N.~D.; Sen,~A.; Velegol,~D. Chemotaxis of
  nonbiological colloidal rods. \emph{Physical Review Letters} \textbf{2007},
  \emph{99}, 178103\relax
\mciteBstWouldAddEndPuncttrue
\mciteSetBstMidEndSepPunct{\mcitedefaultmidpunct}
{\mcitedefaultendpunct}{\mcitedefaultseppunct}\relax
\EndOfBibitem
\bibitem[Pohl and Stark(2014)Pohl, and Stark]{PohlS2014}
Pohl,~O.; Stark,~H. Dynamic clustering and chemotactic collapse of
  self-phoretic active particles. \emph{Physical Review Letters} \textbf{2014},
  \emph{112}, 238303\relax
\mciteBstWouldAddEndPuncttrue
\mciteSetBstMidEndSepPunct{\mcitedefaultmidpunct}
{\mcitedefaultendpunct}{\mcitedefaultseppunct}\relax
\EndOfBibitem
\bibitem[Saha \latin{et~al.}(2014)Saha, Golestanian, and Ramaswamy]{SahaGR2014}
Saha,~S.; Golestanian,~R.; Ramaswamy,~S. Clusters, asters, and collective
  oscillations in chemotactic colloids. \emph{Physical Review E} \textbf{2014},
  \emph{89}, 062316\relax
\mciteBstWouldAddEndPuncttrue
\mciteSetBstMidEndSepPunct{\mcitedefaultmidpunct}
{\mcitedefaultendpunct}{\mcitedefaultseppunct}\relax
\EndOfBibitem
\bibitem[Liebchen \latin{et~al.}(2015)Liebchen, Marenduzzo, Pagonabarraga, and
  Cates]{LiebchenMPC2015}
Liebchen,~B.; Marenduzzo,~D.; Pagonabarraga,~I.; Cates,~M.~E. Clustering and
  pattern formation in chemorepulsive active colloids. \emph{Physical Review
  Letters} \textbf{2015}, \emph{115}, 258301\relax
\mciteBstWouldAddEndPuncttrue
\mciteSetBstMidEndSepPunct{\mcitedefaultmidpunct}
{\mcitedefaultendpunct}{\mcitedefaultseppunct}\relax
\EndOfBibitem
\bibitem[Liebchen \latin{et~al.}(2017)Liebchen, Marenduzzo, and
  Cates]{LiebchenMC2017}
Liebchen,~B.; Marenduzzo,~D.; Cates,~M.~E. Phoretic interactions generically
  induce dynamic clusters and wave patterns in active colloids. \emph{Physical
  Review Letters} \textbf{2017}, \emph{118}, 268001\relax
\mciteBstWouldAddEndPuncttrue
\mciteSetBstMidEndSepPunct{\mcitedefaultmidpunct}
{\mcitedefaultendpunct}{\mcitedefaultseppunct}\relax
\EndOfBibitem
\bibitem[Jin \latin{et~al.}(2017)Jin, Kr{\"u}ger, and Maass]{JinKM2017}
Jin,~C.; Kr{\"u}ger,~C.; Maass,~C.~C. Chemotaxis and autochemotaxis of
  self-propelling droplet swimmers. \emph{Proceedings of the National Academy
  of Sciences U.S.A.} \textbf{2017}, \emph{114}, 5089--5094\relax
\mciteBstWouldAddEndPuncttrue
\mciteSetBstMidEndSepPunct{\mcitedefaultmidpunct}
{\mcitedefaultendpunct}{\mcitedefaultseppunct}\relax
\EndOfBibitem
\bibitem[{Lozano} \latin{et~al.}(2016){Lozano}, {ten Hagen}, {L{\"o}wen}, and
  {Bechinger}]{LozanotHLB2016}
{Lozano},~C.; {ten Hagen},~B.; {L{\"o}wen},~H.; {Bechinger},~C. Phototaxis of
  synthetic microswimmers in optical landscapes. \emph{Nature Communications}
  \textbf{2016}, \emph{7}, 12828\relax
\mciteBstWouldAddEndPuncttrue
\mciteSetBstMidEndSepPunct{\mcitedefaultmidpunct}
{\mcitedefaultendpunct}{\mcitedefaultseppunct}\relax
\EndOfBibitem
\bibitem[Garcia \latin{et~al.}(2013)Garcia, Rafa\"{\i}, and
  Peyla]{GarciaRP2013}
Garcia,~X.; Rafa\"{\i},~S.; Peyla,~P. Light control of the flow of phototactic
  microswimmer suspensions. \emph{Physical Review Letters} \textbf{2013},
  \emph{110}, 138106\relax
\mciteBstWouldAddEndPuncttrue
\mciteSetBstMidEndSepPunct{\mcitedefaultmidpunct}
{\mcitedefaultendpunct}{\mcitedefaultseppunct}\relax
\EndOfBibitem
\bibitem[Geiseler \latin{et~al.}(2016)Geiseler, H\"anggi, Marchesoni, Mulhern,
  and Savel'ev]{GeiselerHMMS2016}
Geiseler,~A.; H\"anggi,~P.; Marchesoni,~F.; Mulhern,~C.; Savel'ev,~S.
  Chemotaxis of artificial microswimmers in active density waves.
  \emph{Physical Review E} \textbf{2016}, \emph{94}, 012613\relax
\mciteBstWouldAddEndPuncttrue
\mciteSetBstMidEndSepPunct{\mcitedefaultmidpunct}
{\mcitedefaultendpunct}{\mcitedefaultseppunct}\relax
\EndOfBibitem
\bibitem[Geiseler \latin{et~al.}(2017)Geiseler, H\"anggi, and
  Marchesoni]{GeiselerHM2017a}
Geiseler,~A.; H\"anggi,~P.; Marchesoni,~F. Self-polarizing microswimmers in
  active density waves. \emph{Scientific Reports} \textbf{2017}, \emph{7},
  41884\relax
\mciteBstWouldAddEndPuncttrue
\mciteSetBstMidEndSepPunct{\mcitedefaultmidpunct}
{\mcitedefaultendpunct}{\mcitedefaultseppunct}\relax
\EndOfBibitem
\bibitem[Geiseler \latin{et~al.}(2017)Geiseler, H\"anggi, and
  Marchesoni]{GeiselerHM2017b}
Geiseler,~A.; H\"anggi,~P.; Marchesoni,~F. Taxis of artificial swimmers in a
  spatio-temporally modulated activation medium. \emph{Entropy} \textbf{2017},
  \emph{19}, 97\relax
\mciteBstWouldAddEndPuncttrue
\mciteSetBstMidEndSepPunct{\mcitedefaultmidpunct}
{\mcitedefaultendpunct}{\mcitedefaultseppunct}\relax
\EndOfBibitem
\bibitem[Sharma and Brader(2017)Sharma, and Brader]{SharmaB2017}
Sharma,~A.; Brader,~J.~M. Brownian systems with spatially inhomogeneous
  activity. \emph{Physical Review E} \textbf{2017}, \emph{96}, 032604\relax
\mciteBstWouldAddEndPuncttrue
\mciteSetBstMidEndSepPunct{\mcitedefaultmidpunct}
{\mcitedefaultendpunct}{\mcitedefaultseppunct}\relax
\EndOfBibitem
\bibitem[{Dai} \latin{et~al.}(2016){Dai}, {Wang}, {Xiong}, {Zhan}, {Dai}, {Li},
  {Feng}, and {Tang}]{DaiWXZDLFT2016}
{Dai},~B.; {Wang},~J.; {Xiong},~Z.; {Zhan},~X.; {Dai},~W.; {Li},~C.-C.;
  {Feng},~S.-P.; {Tang},~J. {Programmable artificial phototactic microswimmer}.
  \emph{Nature Nanotechnology} \textbf{2016}, \emph{11}, 1087--1092\relax
\mciteBstWouldAddEndPuncttrue
\mciteSetBstMidEndSepPunct{\mcitedefaultmidpunct}
{\mcitedefaultendpunct}{\mcitedefaultseppunct}\relax
\EndOfBibitem
\bibitem[Li \latin{et~al.}(2016)Li, Wu, Qin, Zhao, and Liu]{LiWQZL2016}
Li,~W.; Wu,~X.; Qin,~H.; Zhao,~Z.; Liu,~H. Light-driven and light-guided
  microswimmers. \emph{Advanced Functional Materials} \textbf{2016}, \emph{26},
  3164--3171\relax
\mciteBstWouldAddEndPuncttrue
\mciteSetBstMidEndSepPunct{\mcitedefaultmidpunct}
{\mcitedefaultendpunct}{\mcitedefaultseppunct}\relax
\EndOfBibitem
\bibitem[{Lozano} and {Bechinger}(2019){Lozano}, and {Bechinger}]{LozanoB2019}
{Lozano},~C.; {Bechinger},~C. {Diffusing wave paradox of phototactic particles
  in traveling light pulses}. \emph{Nature Communications} \textbf{2019},
  \emph{10}, 2495\relax
\mciteBstWouldAddEndPuncttrue
\mciteSetBstMidEndSepPunct{\mcitedefaultmidpunct}
{\mcitedefaultendpunct}{\mcitedefaultseppunct}\relax
\EndOfBibitem
\bibitem[Lozano \latin{et~al.}(2019)Lozano, Liebchen, {ten Hagen}, Bechinger,
  and L\"owen]{LozanoLtHBL2019}
Lozano,~C.; Liebchen,~B.; {ten Hagen},~B.; Bechinger,~C.; L\"owen,~H.
  Propagating density spikes in light-powered motility-ratchets. \emph{Soft
  Matter} \textbf{2019}, \emph{15}, 5185--5192\relax
\mciteBstWouldAddEndPuncttrue
\mciteSetBstMidEndSepPunct{\mcitedefaultmidpunct}
{\mcitedefaultendpunct}{\mcitedefaultseppunct}\relax
\EndOfBibitem
\bibitem[Palagi \latin{et~al.}(2019)Palagi, Singh, and Fischer]{PalagiSF2019}
Palagi,~S.; Singh,~D.~P.; Fischer,~P. Light-controlled micromotors and soft
  microrobots. \emph{Advanced Optical Materials} \textbf{2019}, \emph{7},
  1900370\relax
\mciteBstWouldAddEndPuncttrue
\mciteSetBstMidEndSepPunct{\mcitedefaultmidpunct}
{\mcitedefaultendpunct}{\mcitedefaultseppunct}\relax
\EndOfBibitem
\bibitem[Ghosh \latin{et~al.}(2015)Ghosh, Li, Marchesoni, and
  Nori]{GhoshLMN2015}
Ghosh,~P.~K.; Li,~Y.; Marchesoni,~F.; Nori,~F. Pseudochemotactic drifts of
  artificial microswimmers. \emph{Physical Review E} \textbf{2015}, \emph{92},
  012114\relax
\mciteBstWouldAddEndPuncttrue
\mciteSetBstMidEndSepPunct{\mcitedefaultmidpunct}
{\mcitedefaultendpunct}{\mcitedefaultseppunct}\relax
\EndOfBibitem
\bibitem[Magiera and Brendel(2015)Magiera, and Brendel]{MagieraB2015}
Magiera,~M.~P.; Brendel,~L. Trapping of interacting propelled colloidal
  particles in inhomogeneous media. \emph{Physical Review E} \textbf{2015},
  \emph{92}, 012304\relax
\mciteBstWouldAddEndPuncttrue
\mciteSetBstMidEndSepPunct{\mcitedefaultmidpunct}
{\mcitedefaultendpunct}{\mcitedefaultseppunct}\relax
\EndOfBibitem
\bibitem[Grauer \latin{et~al.}(2018)Grauer, L\"owen, and Janssen]{GrauerLJ2017}
Grauer,~J.; L\"owen,~H.; Janssen,~L. M.~C. Spontaneous membrane formation and
  self-encapsulation of active rods in an inhomogeneous motility field.
  \emph{Physical Review E} \textbf{2018}, \emph{97}, 022608\relax
\mciteBstWouldAddEndPuncttrue
\mciteSetBstMidEndSepPunct{\mcitedefaultmidpunct}
{\mcitedefaultendpunct}{\mcitedefaultseppunct}\relax
\EndOfBibitem
\bibitem[Scacchi \latin{et~al.}(2019)Scacchi, Brader, and
  Sharma]{ScacchiBS2019}
Scacchi,~A.; Brader,~J.~M.; Sharma,~A. Escape rate of transiently active
  Brownian particle in one dimension. \emph{Physical Review E} \textbf{2019},
  \emph{100}, 012601\relax
\mciteBstWouldAddEndPuncttrue
\mciteSetBstMidEndSepPunct{\mcitedefaultmidpunct}
{\mcitedefaultendpunct}{\mcitedefaultseppunct}\relax
\EndOfBibitem
\bibitem[Vuijk \latin{et~al.}(2019)Vuijk, Brader, and Sharma]{VuijkBS2019}
Vuijk,~H.~D.; Brader,~J.~M.; Sharma,~A. Anomalous fluxes in overdamped Brownian
  dynamics with Lorentz force. \emph{Journal of Statistical Mechanics: Theory
  and Experiment} \textbf{2019}, \emph{2019}, 063203\relax
\mciteBstWouldAddEndPuncttrue
\mciteSetBstMidEndSepPunct{\mcitedefaultmidpunct}
{\mcitedefaultendpunct}{\mcitedefaultseppunct}\relax
\EndOfBibitem
\bibitem[Merlitz \latin{et~al.}(2018)Merlitz, Vuijk, Brader, Sharma, and
  Sommer]{MerlitzVBSS2018}
Merlitz,~H.; Vuijk,~H.~D.; Brader,~J.; Sharma,~A.; Sommer,~J.-U. Linear
  response approach to active Brownian particles in time-varying activity
  fields. \emph{Journal of Chemical Physics} \textbf{2018}, \emph{148},
  194116\relax
\mciteBstWouldAddEndPuncttrue
\mciteSetBstMidEndSepPunct{\mcitedefaultmidpunct}
{\mcitedefaultendpunct}{\mcitedefaultseppunct}\relax
\EndOfBibitem
\bibitem[Liebchen and L\"owen(2019)Liebchen, and L\"owen]{LiebchenL2019}
Liebchen,~B.; L\"owen,~H. Optimal navigation strategies for active particles.
  \emph{EPL} \textbf{2019}, \emph{127}, 34003\relax
\mciteBstWouldAddEndPuncttrue
\mciteSetBstMidEndSepPunct{\mcitedefaultmidpunct}
{\mcitedefaultendpunct}{\mcitedefaultseppunct}\relax
\EndOfBibitem
\bibitem[Maggi \latin{et~al.}(2018)Maggi, Angelani, Frangipane, and
  Di~Leonardo]{MaggiAFDiL2018}
Maggi,~C.; Angelani,~L.; Frangipane,~G.; Di~Leonardo,~R. Currents and
  flux-inversion in photokinetic active particles. \emph{Soft Matter}
  \textbf{2018}, \emph{14}, 4958--4962\relax
\mciteBstWouldAddEndPuncttrue
\mciteSetBstMidEndSepPunct{\mcitedefaultmidpunct}
{\mcitedefaultendpunct}{\mcitedefaultseppunct}\relax
\EndOfBibitem
\bibitem[Uspal(2019)]{Uspal2019}
Uspal,~W.~E. Theory of light-activated catalytic Janus particles. \emph{Journal
  of Chemical Physics} \textbf{2019}, \emph{150}, 114903\relax
\mciteBstWouldAddEndPuncttrue
\mciteSetBstMidEndSepPunct{\mcitedefaultmidpunct}
{\mcitedefaultendpunct}{\mcitedefaultseppunct}\relax
\EndOfBibitem
\bibitem[{Ni} \latin{et~al.}(2016){Ni}, {Leemann}, {Buttinoni}, {Isa}, and
  {Wolf}]{NiLBIW2016}
{Ni},~S.; {Leemann},~J.; {Buttinoni},~I.; {Isa},~L.; {Wolf},~H. {Programmable
  colloidal molecules from sequential capillarity-assisted particle assembly}.
  \emph{Science Advances} \textbf{2016}, \emph{2}, e1501779\relax
\mciteBstWouldAddEndPuncttrue
\mciteSetBstMidEndSepPunct{\mcitedefaultmidpunct}
{\mcitedefaultendpunct}{\mcitedefaultseppunct}\relax
\EndOfBibitem
\bibitem[Ma \latin{et~al.}(2015)Ma, Yang, Zhao, and Wu]{EHDflow}
Ma,~F.; Yang,~X.; Zhao,~H.; Wu,~N. Inducing propulsion of colloidal dimers by
  breaking the symmetry in electrohydrodynamic flow. \emph{Physical Review
  Letters} \textbf{2015}, \emph{115}, 208302\relax
\mciteBstWouldAddEndPuncttrue
\mciteSetBstMidEndSepPunct{\mcitedefaultmidpunct}
{\mcitedefaultendpunct}{\mcitedefaultseppunct}\relax
\EndOfBibitem
\bibitem[Ni \latin{et~al.}(2017)Ni, Marini, Buttinoni, Wolf, and
  Isa]{NiMBWI2017}
Ni,~S.; Marini,~E.; Buttinoni,~I.; Wolf,~H.; Isa,~L. Hybrid colloidal
  microswimmers through sequential capillary assembly. \emph{Soft Matter}
  \textbf{2017}, \emph{13}, 4252--4259\relax
\mciteBstWouldAddEndPuncttrue
\mciteSetBstMidEndSepPunct{\mcitedefaultmidpunct}
{\mcitedefaultendpunct}{\mcitedefaultseppunct}\relax
\EndOfBibitem
\bibitem[{Fernandez-Rodriguez} \latin{et~al.}(2019){Fernandez-Rodriguez},
  {Grillo}, {Alvarez}, {Rathlef}, {Buttinoni}, {Volpe}, and
  {Isa}]{Arxive_magnetic}
{Fernandez-Rodriguez},~M.~A.; {Grillo},~F.; {Alvarez},~L.; {Rathlef},~M.;
  {Buttinoni},~I.; {Volpe},~G.; {Isa},~L. {Active colloids with
  position-dependent rotational diffusivity}. \emph{arXiv:1911.02291,}
  \textbf{2019}, \relax
\mciteBstWouldAddEndPunctfalse
\mciteSetBstMidEndSepPunct{\mcitedefaultmidpunct}
{}{\mcitedefaultseppunct}\relax
\EndOfBibitem
\bibitem[{Hu} \latin{et~al.}(2015){Hu}, {Wysocki}, {Winkler}, and
  {Gompper}]{HuWWG2015}
{Hu},~J.; {Wysocki},~A.; {Winkler},~R.~G.; {Gompper},~G. {Physical sensing of
  surface properties by microswimmers -- directing bacterial motion via wall
  slip}. \emph{Scientific Reports} \textbf{2015}, \emph{5}, 9586\relax
\mciteBstWouldAddEndPuncttrue
\mciteSetBstMidEndSepPunct{\mcitedefaultmidpunct}
{\mcitedefaultendpunct}{\mcitedefaultseppunct}\relax
\EndOfBibitem
\bibitem[Squires and Bazant(2006)Squires, and Bazant]{squires_bazant_2006}
Squires,~T.~M.; Bazant,~M.~Z. Breaking symmetries in induced-charge
  electro-osmosis and electrophoresis. \emph{Journal of Fluid Mechanics}
  \textbf{2006}, \emph{560}, 65--101\relax
\mciteBstWouldAddEndPuncttrue
\mciteSetBstMidEndSepPunct{\mcitedefaultmidpunct}
{\mcitedefaultendpunct}{\mcitedefaultseppunct}\relax
\EndOfBibitem
\bibitem[Gangwal \latin{et~al.}(2008)Gangwal, Cayre, Bazant, and
  Velev]{Gangwal2008}
Gangwal,~S.; Cayre,~O.~J.; Bazant,~M.~Z.; Velev,~O.~D. Induced-Charge
  Electrophoresis of Metallodielectric Particles. \emph{Physical Review
  Letters} \textbf{2008}, \emph{100}, 058302\relax
\mciteBstWouldAddEndPuncttrue
\mciteSetBstMidEndSepPunct{\mcitedefaultmidpunct}
{\mcitedefaultendpunct}{\mcitedefaultseppunct}\relax
\EndOfBibitem
\bibitem[Yan \latin{et~al.}(2016)Yan, Han, Zhang, Xu, Luijten, and
  Granick]{Yan2016}
Yan,~J.; Han,~M.; Zhang,~J.; Xu,~C.; Luijten,~E.; Granick,~S. Reconfiguring
  active particles by electrostatic imbalance. \emph{Nature Materials}
  \textbf{2016}, \emph{15}, 1095--1099\relax
\mciteBstWouldAddEndPuncttrue
\mciteSetBstMidEndSepPunct{\mcitedefaultmidpunct}
{\mcitedefaultendpunct}{\mcitedefaultseppunct}\relax
\EndOfBibitem
\bibitem[Kurzthaler and Franosch(2017)Kurzthaler, and
  Franosch]{KurzthalerF2017}
Kurzthaler,~C.; Franosch,~T. Intermediate scattering function of an anisotropic
  Brownian circle swimmer. \emph{Soft Matter} \textbf{2017}, \emph{13},
  6396--6406\relax
\mciteBstWouldAddEndPuncttrue
\mciteSetBstMidEndSepPunct{\mcitedefaultmidpunct}
{\mcitedefaultendpunct}{\mcitedefaultseppunct}\relax
\EndOfBibitem
\bibitem[Wittkowski and L{\"o}wen(2012)Wittkowski, and
  L{\"o}wen]{WittkowskiL2012}
Wittkowski,~R.; L{\"o}wen,~H. Self-propelled {B}rownian spinning top: dynamics
  of a biaxial swimmer at low {R}eynolds numbers. \emph{Physical Review E}
  \textbf{2012}, \emph{85}, 021406\relax
\mciteBstWouldAddEndPuncttrue
\mciteSetBstMidEndSepPunct{\mcitedefaultmidpunct}
{\mcitedefaultendpunct}{\mcitedefaultseppunct}\relax
\EndOfBibitem
\bibitem[Vo{\ss} and Wittkowski(2018)Vo{\ss}, and Wittkowski]{VossW2018}
Vo{\ss},~J.; Wittkowski,~R. Hydrodynamic resistance matrices of colloidal
  particles with various shapes. \emph{arXiv:1811.01269,} \textbf{2018}, \relax
\mciteBstWouldAddEndPunctfalse
\mciteSetBstMidEndSepPunct{\mcitedefaultmidpunct}
{}{\mcitedefaultseppunct}\relax
\EndOfBibitem
\bibitem[{Cohen} and {Golestanian}(2014){Cohen}, and {Golestanian}]{CohenG2014}
{Cohen},~J.~A.; {Golestanian},~R. {Emergent cometlike swarming of optically
  driven thermally active colloids}. \emph{Physical Review Letters}
  \textbf{2014}, \emph{112}, 068302\relax
\mciteBstWouldAddEndPuncttrue
\mciteSetBstMidEndSepPunct{\mcitedefaultmidpunct}
{\mcitedefaultendpunct}{\mcitedefaultseppunct}\relax
\EndOfBibitem
\end{mcitethebibliography}
\end{document}